\documentclass[english,pra,aps,twocolumn,floatfix,showpacs,tightenlines,superscriptaddress,amsmath,amssymb]{revtex4}
\usepackage[latin1]{inputenc}
\usepackage{verbatim}
\usepackage{graphicx}

\makeatletter

\newcommand{\ket}[1]{|#1\rangle}
\newcommand{\bra}[1]{\langle#1|}

\newcommand{\eq}[1]{Eq.~(\ref{#1})}

\newcommand{\fig}[1]{Fig.~\ref{#1}}

\def\1{1\negthickspace{\rm I}}

\usepackage{babel}
\makeatother
\begin{document}

\title{Long-distance entanglement in many-body atomic and optical systems}

\author{Salvatore M. Giampaolo}

\affiliation{Dipartimento di Matematica e Informatica,
Universit\`{a} degli Studi di Salerno, Via Ponte don Melillo,
I-84084 Fisciano (SA), Italy}

\affiliation{CNR-INFM Coherentia, Napoli, Italy; CNISM Unit\`{a} di
Salerno; and INFN Sezione di Napoli, Gruppo collegato di Salerno,
Baronissi (SA), Italy}

\author{Fabrizio Illuminati}

\affiliation{Dipartimento di Matematica e Informatica,
Universit\`{a} degli Studi di Salerno, Via Ponte don Melillo,
I-84084 Fisciano (SA), Italy}

\affiliation{CNR-INFM Coherentia, Napoli, Italy; CNISM Unit\`{a} di
Salerno; and INFN Sezione di Napoli, Gruppo collegato di Salerno,
Baronissi (SA), Italy}

\affiliation{ISI Foundation for Scientific Interchange, Villa
Gualino, Viale Settimio Severo 65, I-10133 Torino, Italy}

\pacs{03.67.Hk, 03.67.Mn, 75.10.Pq}

\begin{abstract}
We discuss the phenomenon of long-distance entanglement in the
ground state of quantum spin models, its use in high-fidelity and
robust quantum communication, and its realization in many-body
systems of ultracold atoms in optical lattices and in arrays of
coupled optical cavities. We investigate $XX$ quantum spin models on
one-dimensional lattices with open ends and different patterns of
site-dependent interaction couplings, singling out two general
settings: Patterns that allow for perfect long-distance entanglement
(LDE) in the ground state of the system, namely such that the
end-to-end entanglement remains finite in the thermodynamic limit,
and patterns of quasi long-distance entanglement (QLDE) in the
ground state of the system, namely, such such that the end-to-end
entanglement vanishes with a very slow power-law decay as the length
of the spin chain is increased. We discuss physical realizations of
these models in ensembles of ultracold bosonic atoms loaded in
optical lattices. We show how, using either suitably engineered
super-lattice structures or exploiting the presence of edge
impurities in lattices with single periodicity, it is possible to
realize models endowed with nonvanishing LDE or QLDE. We then study
how to realize models that optimize the robustness of QLDE at finite
temperature and in the presence of imperfections using suitably
engineered arrays of coupled optical cavities. For both cases the
numerical estimates of the end-to-end entanglement in the actual
physical systems are thoroughly compared with the analytical results
obtained for the spin model systems. We finally introduce LDE-based
schemes of long-distance quantum teleportation in linear arrays of
coupled cavities and show that they allow for high-fidelity and high
success rates even at moderately high temperatures.
\end{abstract}

\date{November 19, 2009}

\maketitle

\section{Introduction}
\label{introduction} Quantum entanglement plays a crucial role in
many areas of quantum information science \cite{NielsenChuang},
including, among others, quantum cryptography and secure quantum key
distribution \cite{Gisin}, and quantum communication
\cite{Communication}. For this reason, much work has been dedicated
to single out ways to produce useful entanglement for efficient and
robust implementation of quantum information tasks. The most natural
way to create entanglement between two or more constituents of a
quantum information device appears to be by means of direct
interactions because, intuitively, large amounts of entanglement
should be associated to the presence of strong quantum correlations.
However, from a general quantum informatic perspective, an even more
desirable goal is to envisage ways to produce large amounts of
entanglement shared between distant and generally not directly
interacting constituents. Along the way to accomplish this task, a
first step has been realized by introducing the concept of
localizable entanglement \cite{Verstraete}, namely the rate of
entanglement that can be concentrated on a pair of arbitrarily
distant constituents by performing optimal local measurements onto
the remainder of the system. More recently, it has been shown that
the ground state of some spin models with finite correlation length
defined on one-dimensional chains with open ends can support large
values of the end-to-end entanglement between the initial and final
points of the chain, insensitive or only very weakly sensitive to
the size of the system\cite{Bologna1,Bologna2,Salerno1,Salerno2}.
This type of entanglement has thus been dubbed Long-Distance
Entanglement (LDE). In this approach, the guiding principle is to
look for systems whose GS can support {\it intrinsically} large
amounts of (long-distance) entanglement, even at moderately high
temperatures and in the presence of bulk imperfections, without the
need for performing operations and measurements or for continuous
dynamical controls and adjustments of the couplings. Clearly, the
property of LDE would be very appealing in the quest to design
quantum information devices, that perform tasks efficiently and at
the same time are sufficiently robust against decoherence, based on
(properly engineered) many-body systems of condensed matter that
could possibly be realized with currently available technologies, or
technologies that may be in view in the near future.

The scope of the present work is to investigate various theoretical
aspects of the phenomenon of LDE in quantum many-body physics, to
introduce possible concrete experimental implementations with
currently available many-body systems of atomic physics and quantum
optics, and to discuss simple schemes of LDE-based efficient and
robust quantum communication. Concerning the theoretical modeling,
here and in the following we will consider as paradigmatic test-beds
some very simple isotropic quantum spin models on open chains with
$XX$-type interactions of site-dependent strength. For suitable
choices of the sets of site-dependent couplings, these models can
show large values of the entanglement shared by the end spins of the
chain \cite{Salerno1,Salerno2}. Notwithstanding their apparent
simplicity, they are a very powerful and useful tool in order to
investigate, identify, classify, and summarize the different
possible types of LDE, and the properties associated to different
patterns of site-dependent interaction strengths. Moreover, these
different classes of $XX$ quantum spin models with site-dependent
interaction couplings turn out to be very well suited to assess the
experimentally relevant problem of the vanishing of the energy gap
as a function of the size of the system, and to identify,
unambiguously, the optimal range of parameters compatible with the
largest achievable value of LDE at fixed values of the temperature
and of the length of the chain.

After establishing the general theoretical framework, we will
analyze some possible realizations based on two many-body systems of
current experimental interest and that are moreover evolving and
being developed at a fast pace, namely systems of ultracold neutral
atoms in optical lattices \cite{Bloch} and arrays of coupled
cavities \cite{Plenio1}. These two systems share some common
features. They can be seen as ensembles of local structures that
interact with each other via the exchange of bosonic particles,
atoms in the case of optical lattices, and photons in the case of
coupled cavities. Both systems allow to simulate quantum spin models
by effectively reducing the dimension of the local state space, and
by realizing spin-spin interactions via the inter-site hopping
amplitudes. However, while the hopping amplitude in an array of
coupled optical microcavities can be tuned at will, at least in
principle, either by adjusting the inter-cavity overlaps and/or by
tuning the cavities' fundamental physical parameters (Rabi
couplings, quality factors, etc.) of every single cavity
(single-site addressing), this task appears to be much more
challenging in optical lattices, for which the inter-site properties
are indissolubly woven with the frequency of the external laser
potentials, and hence site-dependent interactions between the
effective spins can be engineered either by introducing atomic
impurities and/or by designing super-lattice structures. We will
discuss merits and disadvantages of both types of systems in the
quest for the experimental demonstration of LDE and its use in the
realization of efficient, robust, and high-fidelity quantum
information tasks.

The paper is organized as follows: In Section II we introduce the
general $XX$ spin chain Hamiltonian with arbitrary site-dependent
couplings and review the methods to determine its spectrum, its
ground-state properties, and the end-to-end concurrence. In Section
III we investigate the properties of LDE for different patterns of
site-dependent interaction couplings, singling out two different
general situations: Patterns that allow for perfect LDE, namely such
that the end-to-end entanglement remains finite in the limit of
diverging size of the system, and quasi-LDE (or imperfect LDE), i.e.
such that the end-to-end entanglement vanishes with a very slow
power-law decay as the length of the chain increases. In Section IV
we discuss how to realize or, better, how to simulate the models
analyzed in Section III with ensembles of ultracold bosonic atoms
loaded in optical lattices. In the literature there exist several
proposals to implement LDE-free spin models, either with uniform
\cite{Sorensen,Duan,Kuklov} or random \cite{Sanpera} spin-spin
couplings, using optical lattices in various settings. Here we show
how, using either suitably engineered super-lattice structures or
exploiting the presence of edge impurities in lattices with single
periodicity, it is possible to realize models endowed with
nonvanishing LDE in the ground state. In section V we investigate
how to realize spin models supporting LDE by the use of suitable
array structures of coupled cavities. In fact, these systems are
being intensively studied in relation to their ability to
realize/simulate collective phenomena typical of strongly correlated
systems of condensed matter
\cite{Hartmann,Greentree,Angelakis,Rossini,Hartmann-2,Plenio1}.
Concerning the simulation of spin models, we show how, at least in
principle, this type of system allows to simulate straightforwardly
$XX$ models with arbitrary patterns of site-dependent couplings and
different types of LDE. In both sections numerical estimates of the
end-to-end entanglement in the actual physical systems are
thoroughly compared with the results obtained in the simulated spin
model systems. In Section VI we discuss the implementation of an
LDE-based scheme of long-distance quantum teleportation exploiting
open linear arrays of coupled cavities. In particular, we show that
this scheme allows for high-fidelity teleportation even at
moderately high temperatures and in the presence of noise. Finally,
in the conclusions we summarize our findings and discuss some
outlooks on possible future developments along this line of
research.

\section{General structure of $XX$ quantum spin models on open chains}
\label{generalapproach}

As anticipated in the introduction, in this Section we discuss the
properties of $XX$ quantum spin-$1/2$ models defined on
one-dimensional lattices with open ends, with specific patterns of
nearest-neighbor interactions. Such models are all special instances
of the general $XX$ Hamiltonian
\begin{equation}
H_{XX} = \sum_{i=1}^{N-1}J_{i}\left(S_{i}^{x}S_{i+1}^{x} +
S_{i}^{y}S_{i+1}^{y}\right)\; , \label{HamiltonianaGenerica}
\end{equation}
where $J_{i}$ is the interaction strength between spins at nearest
neighboring sites $i$ and $i+1$, $S_{i}^{\alpha}$ is the spin-$1/2$
operator defined at site $i$, and $N$ is the total number of sites
(spins) or, equivalently, the length of the chain. The spectrum of
this Hamiltonian can be determined exactly by a straightforward
generalization of the methods first discussed by Lieb, Schultz, and
Mattis \cite{Lieb}. The first step in the procedure is to perform a
Jordan-Wigner transformation \cite{JordanWigner},
\begin{eqnarray}
& &
S_{i}^{+}=c_{i}^{\dagger}e^{i\pi\sum_{j=1}^{i-1}c_{j}^{\dagger}c_{j}}\;,\qquad
S_{i}^{-}
=e^{-i\pi\sum_{j=1}^{i-1}c_{j}^{\dagger}c_{j}}c_{i}\;,\nonumber \\
& & S_{i}^{z}=c_{i}^{\dagger}c_{i}-\frac{\1}{2}\;,
\end{eqnarray}
where $S_{j}^{\pm}=S_{j}^{x}\pm iS_{j}^{y}$. As a result, the
Hamiltonian (\ref{HamiltonianaGenerica}) is mapped in the free
fermion Hamiltonian
\begin{equation}
H=\frac{1}{2}\sum_{i=1}^{N-1}J_{i}\left(c_{i}^{\dagger}c_{i+1}+
c_{i+1}^{\dagger}c_{i}\right)=\mathbf{c}^{\dagger}M\mathbf{c}\;,
\label{Trasformata}
\end{equation}
where
$\mathbf{c}^{\dagger}=\left(c_{1}^{\dagger},\ldots,c_{N}^{\dagger}\right)$
($\mathbf{c}$) is the vector of the $N$ fermionic creation
(annihilation) operators, one for each site of the lattice, and the
adjacency matrix $M$ reads
\begin{equation}
M=\frac{1}{2}\left|\begin{array}{cccccc}
0 & J_{1} & 0 & \cdots &  & 0\\
J_{1} & 0 & J_{2}\\
0 & J_{2} & 0 &  &  & \vdots\\
\vdots &  &  & \ddots & J_{N-2} & 0\\
 &  &  & J_{N-2} & 0 & J_{N-1}\\
0 &  & \cdots & 0 & J_{N-1} & 0\end{array}\right|\;.
\label{Mgenerica}
\end{equation}
Given the GS of the system, we are interested in the evaluation of
the end-to-end entanglement in the reduced two-qubit state between
the first spin at site 1 and the last spin at site $N$. One thus
needs to determine the spin-spin concurrence (entanglement of
formation) between the end points of the chain. This quantity can be
computed exactly for any two-qubit state (pure or mixed), thanks to
the celebrated Wootters formula \cite{Wootters}, and the task is
left to obtain its explicit expression in the reduced state of the
two end-point spins. To this purpose, we need to calculate
explicitly all the possible forms of two-point correlations in the
GS. The Hamiltonian (\ref{HamiltonianaGenerica}) is symmetric under
rotations of the spins around the $z$-axis, so that the only
nonvanishing correlations are $\langle
S_{i}^{x}S_{j}^{x}\rangle=\langle S_{i}^{y}S_{j}^{y}\rangle$,
$\langle S_{i}^{z}S_{j}^{z}\rangle$ and $\langle S_{i}^{z}\rangle$.
In the absence of external magnetic fields, $\pi$-rotations around
the $x$ and $y$ axes are symmetries of the model, which additionally
implies $\langle S_{i}^{z}\rangle=0$ at every site. Thanks to the
aforementioned symmetries, the two-point reduced density matrix
$\rho_{i,j}$, obtained by tracing the ground-state density matrix of
the entire system over all spins except the ones at sites $\{
i,j\}$, has the form
\begin{equation}
\rho_{i,j}=\frac{\1}{4}+\langle
S_{i}^{x}S_{j}^{x}\rangle\left(\sigma^{x}\otimes\sigma^{x} +
\sigma^{y}\otimes\sigma^{y}\right)+\langle
S_{i}^{z}S_{j}^{z}\rangle\sigma^{z}\otimes\sigma^{z},
\label{ReducedDensity}
\end{equation}
where $\sigma^{x,y,z}$ are the Pauli matrices and
$\langle\cdot\rangle$ is the GS average at temperature $T=0$, or the
thermal one at finite temperature $\beta=(k_{B}T)^{-1}$ with respect
to the Gibbs state $\rho=e^{-\beta H}Z^{-1}$. We are interested in
the case in which $i=1$ and $j=N$ are the two end points of the
chain. In this instance, we have
\begin{eqnarray}
& & S_{1}^{+}S_{N}^{-}+S_{1}^{-}S_{N}^{+}=
-e^{i\pi\mathcal{N}}\left(c_{1}^{\dagger}c_{N}+
c_{N}^{\dagger}c_{1}\right)\;,\nonumber \\
& & S_{1}^{z}S_{N}^{z}=\left(c_{1}^{\dagger}
c_{1}-\frac{1}{2}\right)\left(c_{N}^{\dagger}c_{N}
-\frac{1}{2}\right)\;, \label{Correlazionigeneriche}
\end{eqnarray}
where $\mathcal{N}=\sum_{i=1}^{N}c_{i}^{\dagger}c_{i}$ is the total
number operator. Applying Wick's theorem and taking into account
that $\langle c_{i}^{\dagger}c_{i}\rangle=1/2$, we obtain
\begin{eqnarray}
& & \langle S_{1}^{+}S_{N}^{-}+S_{1}^{-}S_{N}^{+}\rangle= -e^{i\pi
N/2}\left(\langle c_{1}^{\dagger}c_{N}\rangle
+ \langle c_{N}^{\dagger}c_{1}\rangle\right)\;,\nonumber \\
&  & \langle S_{1}^{z}S_{N}^{z}\rangle=-\langle
c_{1}^{\dagger}c_{N}\rangle\langle
c_{N}^{\dagger}c_{1}\rangle.
\end{eqnarray}
Setting $x\equiv\langle
c_{1}^{\dagger}c_{N}\rangle$  we see that the end-to-end reduced
density matrix depends uniquely on this parameter. For reduced
states of the form (\ref{ReducedDensity}) the end to end
concurrence $C_{1,N}$ is readily computed \cite{Zanardi02}.
In our case, we have
\begin{equation}
C_{1,N}=2\max\left\{ 0,\, x^{2}+\left|x\right|-\frac{1}{4}\right\}
\; . \label{equazioneconcurrence}
\end{equation}
The concurrence is thus nonvanishing for
$|x|>(\sqrt{2}-1)/2 \simeq 0.207$, and it reaches the maximum value
$C_{1,N}=1$ for $|x| \rightarrow 1/2$.


It is natural to expect that the existence of a strong quantum
correlation between the two end spins of the chain can be
conveniently exploited for performing tasks in quantum information,
in particular considering teleportation schemes. In the standard
quantum teleportation protocol, two parties A and B share a
maximally entangled state (Bell state). Party A holds also a third
qubit, whose unknown state is to be teleported. If the two end
points of our $XX$ chain share a highly entangled state, that in
some limit may even be asymptotically close to a Bell state, they
can be identified as the two parties, sender and receiver, for a
long-distance, high-fidelity teleportation protocol. The efficiency
of a quantum channel in teleporting an unknown state is quantified
by the fidelity $f$ between the output and the input states,
averaged over all input realizations. The fidelity depends on the
actual properties of the entangled resource $\rho_{1,L}$ (Cfr.
\eq{ReducedDensity}) shared by the end spins of the chain. In fact,
it has been demonstrated that the optimal fidelity depends only on
the {}``fully entangled fraction'' $F_{full}$, according to the
formula $f=(2F_{full}+1)/3$ \cite{Horodecki}. The fully entangled
fraction is defined as the fidelity between the resource
$\rho_{1,L}$ and a maximally entangled state, maximized over all
possible maximally entangled states. For states of the form
\eq{ReducedDensity} it can be easily computed, and reads
$F_{full}=\frac{1}{4}+|x|+x^{2}$ \cite{Badziag}. The associated
teleportation fidelity is thus
\begin{equation}
f=\frac{2\left(\frac{1}{4}+|x|+x^{2}\right)+1}{3} \; ,
\label{generalfidelity}
\end{equation}
and, taking into account \eq{equazioneconcurrence}, from
\eq{generalfidelity} we obtain the expression of the
concurrence $C_{1,N}$ as a function of the fidelity:
\begin{equation}
C_{1,N}=2\max\left\{ 0,\, \frac{3}{2} f-1 \right\} \; .
\label{concurrencefidelity}
\end{equation}
This expression highlights the crucial interplay between
entanglement and efficiency in quantum information protocols. In
fact, due to the high symmetry of states of the form
\eq{ReducedDensity}, a nonvanishing entanglement implies a
nonclassical teleportation fidelity exceeding the classical
threshold $2/3$, and viceversa. Moreover the relation between the
maximum fidelity that can be obtained in a quantum teleportation
between the end points of the spin chain and the end-to-end
\eq{concurrencefidelity} provides a simple route to probe,
via experimental implementations of quantum teleportation
protocols, the actual values of the generated long-distance
entanglement $C_{1,N}$.


All the physical information about model (\ref{Trasformata}) can now
be obtained by diagonalizing the one-body matrix $M$. Let $\xi_{k}$
be the eigenvector of $M$ with eigenvalue $\Lambda_{k}$, where $k$
is a quasi-momentum label. Passing to new fermionic operators via
the transformation $c_{i}=\sum_{k}\xi_{k}^{\left(i\right)}c_{k}$,
the Hamiltonian takes the form
\begin{equation}
H_{XX} = \sum_{k}\Lambda_{k}c_{k}^{\dagger}c_{k}\;.
\end{equation}
The evaluation of the two-point correlation $x$ is finally
straightforward:
\begin{eqnarray}
x & = & \sum_{k,q}\xi_{k}^{\left(1\right)}\xi_{q}^{\left(N\right)}
\langle
c_{k}^{\dagger}c_{q} \rangle \nonumber \\
& & \nonumber \\
& = & \left\{ \begin{array}{ll}
\sum_{\Lambda_{k}<0}\xi_{k}^{\left(1\right)} \xi_{k}^{\left(N\right)} & \mbox{for }T=0 \\
\rule{0mm}{5mm}\sum_{k}\xi_{k}^{\left(1\right)}
\xi_{k}^{\left(N\right)}\frac{1}{1+e^{\beta\Lambda_{k}}} & \mbox{for
}T>0 \; , \end{array} \right. \label{correlationgeneral}
\end{eqnarray}
where $x$ depends on the set of the couplings $J_i$ as well as
on the temperature $T$, whenever $T \neq 0$.

In conclusion, the full evaluation of the end-to-end entanglement in
$XX$ quantum spin models defined on open chains, with arbitrary
patterns of site-dependent nearest-neighbor interactions, is
strictly associated to the complete diagonalization of the $M$
matrix in Eq.(\ref{Mgenerica}). While in some particular cases it is
possible to find the exact analytic expressions of the eigenvalues
and of the eigenvectors of $M$ \cite{Woj05,Salerno1}, in the most
general case the problem has to be solved by exact numerical
diagonalization.

\section{End-to-End entanglement properties of $XX$ spin chains}
\label{Spinresult}

Having reviewed the basic definitions and mathematical tools needed
to analyze the end-to-end entanglement properties in $XX$ spin
chains with arbitrary site-dependent interaction strengths, we need
to look for those sets of couplings that allow for large values of
LDE, possibly robust against thermal decoherence and bulk
imperfections. the required entanglement properties. The basic idea
for concentrating a large amount of entanglement on the two end
spins of the chain is inspired by the observation that for some
frustration-free systems, like the antiferromagnetic Heisenberg
model \cite{Bologna1,Bologna2}, and some $XX$-type models
\cite{Salerno1,Salerno2}, the total energy is minimized by states
tending to form a global singlet. This fact, together with the
property of monogamy for bipartite entanglement \cite{Coffman}, is
the primary cause for the phenomenon of LDE. Namely, if one selects
a set of interaction couplings in the Hamiltonian
Eq.(\ref{HamiltonianaGenerica}) that are such to forbid the end
spins to entangle with any other constituents, then, by monogamy,
strong quantum correlations will develop between them.

\subsection{Perfect Long Distance Entanglement (LDE)}
\label{LDE}

In analogy with the case of the dimerized antiferromagnetic
Heisenberg chain, that was the system in which the phenomenon of LDE
was first discovered \cite{Bologna1}, we begin by reviewing the
properties of the $XX$ spin chain with bonds of alternating
strengths, i.e. a model in which each weak bond ($J_{k-1} = \lambda
J$, with $\lambda << 1$) between neighboring spins at sites is
followed by a much stronger one ($J_{k} = J$) between the successive
pair of neighboring spins. This model has been introduced and
investigated in Ref. \cite{Salerno1}. If the chain is formed of an
even number of sites and the two end points are coupled to their
nearest neighbors with a weak bond, this model fulfills exactly the
requirements needed in order to establish perfect LDE between the
end spins of the chain. Namely, any spin in the chain but the end
ones is subject to a weak (left) interaction and to a strong (right)
interaction. Hence, every spin in the bulk tends to be maximally
entangled with its strongly interacting neighbor. Each of the two
end spins is subject only to a weak bond interaction, and is
therefore excluded from the arrangement in pairs of nearest-neighbor
singlets due to the monogamy property of bipartite entanglement
\cite{Coffman}. Furthermore, because the GS of the model tends to be
a collection of singlets, the two end spin spins are effectively
coupled and develop a nonvanishing LDE mediated by the remainder of
the chain, notwithstanding the fact that they do not interact
directly in any way.

\begin{figure}
\begin{centering} \includegraphics[width=75mm,keepaspectratio]{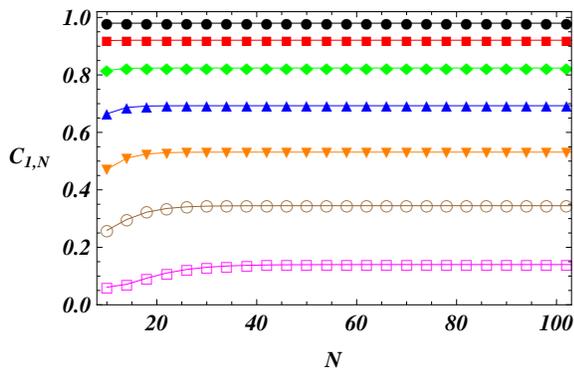}
\par\end{centering}
\caption{End-to-end concurrence $C_{1,N}$ for the 1-D $XX$ spin
model with alternating couplings described by the Hamiltonian
Eq.\ref{LDEHamiltonian}, plotted as a function of the number of
lattice sites $N$ (length of the chain), for different value of the
weak coupling $\lambda$. From bottom to top: magenta empty squares:
$\lambda=0.7$; brown empty circles: $\lambda=0.6$; orange inverted
triangles: $\lambda=0.5$; blue triangles: $\lambda=0.4$; green
diamonds: $\lambda=0.3$; red full squares: $\lambda=0.2$; black full
circles: $\lambda=0.1$.} \label{fig:njp1}
\end{figure}

The above qualitative reasoning can be made precise and
quantitative. The Hamiltonian for the $XX$ spin model with
alternating weak and strong interactions is
\begin{equation}
H = \frac{J}{2} \sum_{i=1}^{N-1}
[(1+\lambda)+(-1)^i(1-\lambda)]\left(S_{i}^{x}S_{i+1}^{x} +
S_{i}^{y}S_{i+1}^{y}\right) \; .
\label{LDEHamiltonian}
\end{equation}
Following the approach described in the previous section, one can
determine the end-to-end concurrence for different values of
$\lambda$ and different lengths of the chain \cite{Salerno1}, as
shown in \fig{fig:njp1}.
We can see that as the weak coupling $\lambda$ decrease below a
threshold value $\lambda_c = 0.765$ \cite{Salerno1}, a nonvanishing
concurrence begins to develop. For a fixed length of the chain,
decreasing $\lambda$ still further (and thus increasing the
difference between the alternating couplings) allows the creation of
dimers between pairs of strongly interacting neighboring spins. The
dimers, in turn, are very weakly coupled to each other. Hence,
because the two end spins of the chain interact very weakly with
their nearest neighbors, dimerization and monogamy of entanglement
force the creation of a strong quantum correlation between the end
points, in analogy with what happens in the Heisenberg case
\cite{Bologna1}. On the other hand, at fixed $\lambda$, the
end-to-end concurrence $C_{1,N}$ as a function of the size $N$ of
the chain converges very rapidly to its thermodynamic-limit
saturation value:
\begin{equation}
C_{1,\infty} = 2\max\left\{
0,\frac{1}{2}-\lambda^{2}+\frac{\lambda^{4}}{4} \right\} \; .
\label{formulefinaliLDE}
\end{equation}

\begin{figure}
\begin{centering} \includegraphics[width=75mm,keepaspectratio]{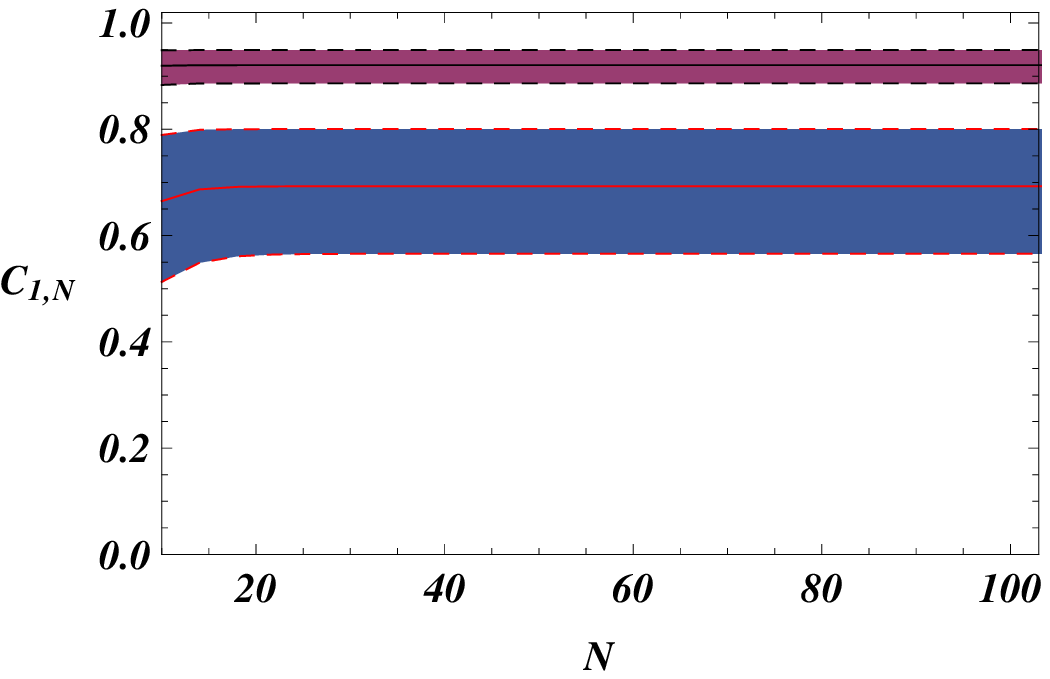}
\par\end{centering}
\caption{End-to-end entanglement as a function of the length of the
chain for the model with noisy couplings described by \eq{LDEHamiltonian-noise},
where $\chi_i$ is a random variable uniformly distributed in the interval
$\{-\bar{\chi},\bar{\chi}\}$, and $\bar{\chi}=0.2$. The violet
area corresponding to $\lambda=0.2$, and the blue one corresponding
to $\lambda=0.4$, are the domains in which the different random samples may fall.
They have been obtained with $10^{4}$ independent samples for each value $N$
of the length of the chain.} \label{fig:njp1-err}
\end{figure}


Next, it is important to verify that the phenomenon of
strong end to end entanglement survives in the
presence of disorder. Indeed, in \fig{fig:njp1-err} we
represent the behavior of the end-to-end entanglement
for different sample realizations of the following
Hamiltonian with disordered couplings:
\begin{equation}
H = \frac{J}{2} \sum_{i=1}^{N-1}
[(1+\lambda)+(-1)^i(1-\lambda)](1+\chi_i) \left(S_{i}^{x}S_{i+1}^{x}
+ S_{i}^{y}S_{i+1}^{y}\right) \; , \label{LDEHamiltonian-noise}
\end{equation}
where $\chi_i$ is a random variable uniformly distributed in the
interval $\{-\bar{\chi},\bar{\chi}\} $. As we may note, the presence
of noise affects bi-directionally the entanglement properties of the
system, and the end to end concurrence is nonvanishing even in the presence
of a noise with a relative weight of $20 \%$. For nearly half of
the samples, noise increases the end to end concurrence.
Moreover, \fig{fig:njp1-err} shows that the stronger the
end-to-end entanglement without noise, the weaker the relative
effect of the noise.


\begin{figure}
\begin{centering} \includegraphics[width=75mm,keepaspectratio]{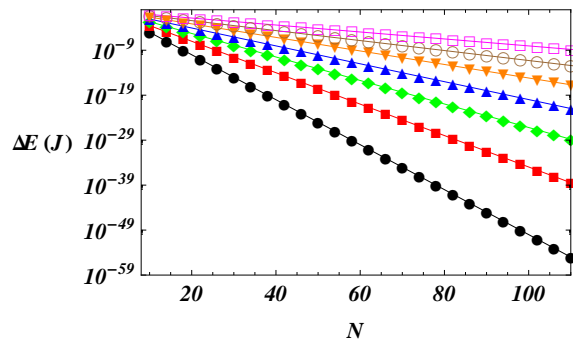}
\par\end{centering}
\caption{Energy Gap $\Delta E (J)$ between the ground and the first
excited state, in units of the coupling energy $J$, for the 1-D $XX$
spin model with alternating couplings described by the Hamiltonian
Eq. (\ref{LDEHamiltonian}), plotted as a function of the length $N$
of the chain for different values of the weak coupling $\lambda$.
From top to bottom: magenta empty squares: $\lambda=0.7$; brown
empty circles: $\lambda=0.6$; orange inverted triangles:
$\lambda=0.5$; blue triangles: $\lambda=0.4$; green diamonds:
$\lambda=0.3$; red full squares: $\lambda=0.2$; black full circles:
$\lambda=0.1$.} \label{fig:njp2}
\end{figure}

It would then seem that in order to obtain a physical system with
strong LDE it should be sufficient to engineer some concrete device
that in appropriate limits simulates/realizes an $XX$ spin chain
chain with alternating weak and strong couplings. Unfortunately,
this is not the case in realistic situations. As shown in Fig.
\ref{fig:njp2}, the phenomenon of perfect LDE is strictly associated
to an energy gap between ground and first excited states that
vanishes exponentially fast as a function of the length of the
chain. Therefore, as soon as the system is at temperatures
comparable with the gap $\Delta E$, the equilibrium Gibbs state
results in an incoherent superposition of singlet and triplet states
in which the LDE vanishes \cite{Bologna2,Salerno1}. Consequently, in
order to realize a nonvanishing entanglement between the two end
points, even in short chains, it would be necessary to reach
temperatures fantastically close to absolute zero, something that is
well beyond reach in any foreseeable future, and in any case of no
use for practical applications.

\begin{figure}
\begin{centering} \includegraphics[width=75mm,keepaspectratio]{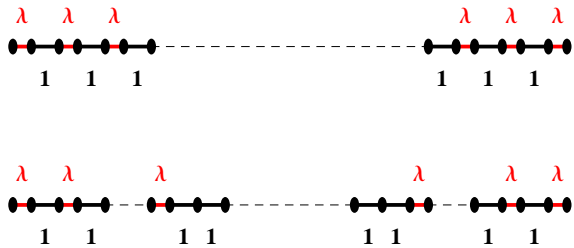}
\par\end{centering}
\caption{Schematic drawing of $XX$ spin chains endowed with perfect
LDE in the GS. Upper drawing: Model Eq.(\ref{LDEHamiltonian}), with
alternating weak ($\lambda <1$) and strong couplings along the
entire chain. Lower drawing: Model Eq.(\ref{LDE-1-Hamiltonian}),
with left and right end regions of alternating weak and strong
couplings and a central (bulk) region of uniform strong couplings.}
\label{fig:ModelliLDE}
\end{figure}

\begin{figure}
\begin{centering} \includegraphics[width=75mm,keepaspectratio]{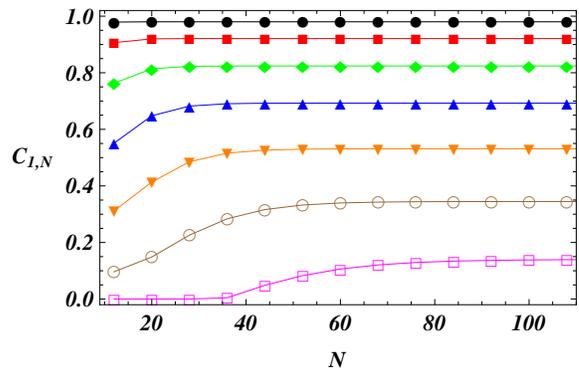}
\par\end{centering}
\caption{End-to-end concurrence $C_{1,N}$ for an $XX$ spin chain
with alternating couplings and a central region with uniform
interactions, described by the Hamiltonian
Eq.(\ref{LDE-1-Hamiltonian}), plotted as a function of the length
$N$ of the chain, for different values of the weak coupling
$\lambda$ and a fixed value of the ratio $\tilde{N}/N=1/4$. From
bottom to top: magenta empty squares: $\lambda=0.7$; brown empty
circles: $\lambda=0.6$; orange inverted triangles: $\lambda=0.5$;
blue triangles: $\lambda=0.4$; green diamonds: $\lambda=0.3$; red
full squares: $\lambda=0.2$; black full circles: $\lambda=0.1$.}
\label{fig:njp3}
\end{figure}

\begin{figure}
\begin{centering} \includegraphics[width=75mm,keepaspectratio]{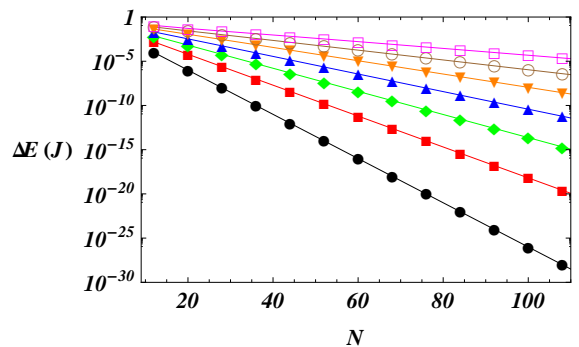}
\par\end{centering}
\caption{Energy gap, in units of the coupling energy $J$, between
the ground and the first exited state for an $XX$ spin chain with
alternating couplings and a central region with uniform
interactions, described by the Hamiltonian
Eq.(\ref{LDE-1-Hamiltonian}). The gap $\Delta E (J)$ is plotted as a
function of the length $N$ of the chain, for different values of the
weak coupling $\lambda$ and at a fixed value of the ratio
$\tilde{N}/N=1/4$. From top to bottom: magenta empty squares:
$\lambda=0.7$; brown empty circles: $\lambda=0.6$; orange inverted
triangles: $\lambda=0.5$; blue triangles: $\lambda=0.4$; green
diamonds: $\lambda=0.3$; red full squares: $\lambda=0.2$; black full
circles: $\lambda=0.1$.} \label{fig:njp4}
\end{figure}

One might still consider realizing the request of perfect LDE by
looking at modifications or alternatives to the pattern of perfectly
alternating weak and strong couplings. Such a pattern has been
introduced with the aim of suppressing the formation of sizeable
entanglement between the end points and the remaining spins of the
lattice. However, it is quite clear that this goal can be achieved
considering different spatial interaction patterns that have the
effect of isolating the bulk of the chain from the end points. This
observation motivates the introduction of the following class of
Hamiltonians:
\begin{eqnarray}
\label{LDE-1-Hamiltonian}
&& H =  \frac{J}{2} \left[\sum_{i=1}^{\tilde{N}-1}
[(1+\lambda)+(-1)^i(1-\lambda)]\left(S_{i}^{x}S_{i+1}^{x}
\right. \right.  \nonumber \\
&& \left. + S_{i}^{y}S_{i+1}^{y}\right) + \sum_{i=\tilde{N}}^{N-\tilde{N}-1}
2\left(S_{i}^{x}S_{i+1}^{x} +
S_{i}^{y}S_{i+1}^{y} \right)  \nonumber \\
&& \left. + \sum_{i=N-\tilde{N}}^{N-1}
[(1+\lambda)+(-1)^i(1-\lambda)]\left(S_{i}^{x}S_{i+1}^{x} +
S_{i}^{y}S_{i+1}^{y}\right) \right] . \nonumber \\
&&
\end{eqnarray}
Models Eq.(\ref{LDEHamiltonian}) and Eq.(\ref{LDE-1-Hamiltonian})
with their spatial patterns of site-dependent couplings are sketched
pictorially in Fig. \ref{fig:ModelliLDE}.
The choice of the spatial pattern of the couplings that
characterizes model Eq.(\ref{LDEHamiltonian}) allows to preserve
perfect LDE between the end points of the $XX$ spin chain, at the
same time drastically increasing the energy gap between the GS and
the first excited states. In practice, it amounts to replace in the
bulk of the chain, i.e. between a reference $\tilde{N}$-th spin and
the $N-\tilde{N}$-th one, the alternating pattern of weak and strong
couplings with a uniform nearest-neighbor interaction. The results
for the end-to-end entanglement as a function of the length of the
chain, for different values of $\lambda$ and at a fixed value of the
ratio $\tilde{N}/N$ are reported in Fig. \ref{fig:njp3}.
In the thermodynamic limit, the model described by Eq.
(\ref{LDE-1-Hamiltonian}) exhibits a LDE behaviour analogous to that
of the model with fully alternating weak and strong couplings
described by Eq. (\ref{LDEHamiltonian}), when the size of the chain
is increased at a fixed, constant value of the ratio $\tilde{N}/N$.
The only difference, as can be seen by comparing Fig. \ref{fig:njp1}
with Fig. \ref{fig:njp3}, is that in the second case the end-to-end
concurrence reaches the saturation value more slowly, with a speed
that decreases when decreasing the ratio $\tilde{N}/N$. While the
essential features of the LDE behaviour are shared also by the
modified model with uniform central interactions, the behaviour of
the energy gap between the ground and the first excited states is
instead strongly affected, as shown in Fig. \ref{fig:njp4}.
As we can see from Fig. \ref{fig:njp4}, the energy gap again
decreases as the length of the chain is increased, as in the
previous case, and it attains values that forbid possible
experimental implementations of LDE at finite temperature (although
higher than the corresponding ones for the model with perfectly
alternating couplings).
Comparing Fig. \ref{fig:njp4} with Fig. \ref{fig:njp2} and taking
into account that in the present case $\tilde{N}=N/4$ the behavior
of the energy gap does not depend on the total length $N$ of the
chain, but rather only on the length $2 \tilde{N} = N/2$ of that part of
the chain characterized by alternating pairings. This dependence of
the energy gap on the length of that part of the chain endowed with
alternate couplings is confirmed by the analysis of systems defined
at different values of $\tilde{N}$.
This fact suggests to investigate further possibilities. Indeed,
one may think of bringing the modification implemented in
Eq.(\ref{LDE-1-Hamiltonian}) to the extreme, in order to realize non
negligible values of LDE at small, but experimentally feasible
temperatures.

\subsection{Quasi Long Distance Entanglement (QLDE)}
\label{QLDE}

\begin{figure}
\begin{centering} \includegraphics[width=75mm,keepaspectratio]{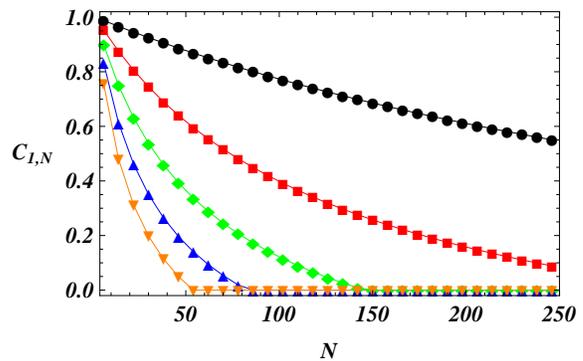}
\par\end{centering}
\caption{End-to-end concurrence $C_{1,N}$ for an $XX$ spin chain
with uniform bulk couplings and two weak end probes described by the
Hamiltonian Eq.(\ref{QLDEHamiltoniana}), plotted as a function of
the length $N$ of the chain for different values of the weak
coupling $\lambda$. From bottom to top: orange inverted triangles:
$\lambda=0.2$; blue triangles: $\lambda=0.16$; green diamonds:
$\lambda=0.12$; red full squares: $\lambda=0.08$; black full
circles: $\lambda=0.04$.} \label{fig:njp5}
\end{figure}

As we have seen at the end of the previous subsection, reducing the
portion of an $XX$ chain with alternating patterns of interaction
does not reduce the entanglement shared by the end points and
increases by several orders of magnitude the energy gap between the
ground and the first excited states, yet, unfortunately, it does not
allow LDE to survive except at unrealistically low temperatures.
Therefore, in order to realize a nonvanishing LDE at low, but
realistically attainable temperatures, we consider taking the limit
of Eq.(\ref{LDE-1-Hamiltonian}) to a model with uniform
nearest-neighbor interactions for all pairs of spins but for the two
end-points, that are connected to the rest of the chain with a weak
bond. Indeed, spin systems allowing for strong end-to-end
correlations should be characterized by interactions between the end
points and their nearest neighbors that are always smaller than the
interactions in the bulk of the chain. Otherwise, if the system does
not meet this criterion, the end points would become strongly
entangled with their neighbors, excluding, due to the monogamy
constraints \cite{Coffman}, the possibility of LDE. Hence, we
consider an open $XX$ spin chain formed by $N-2$ spins with uniform
coupling strengths, plus two weakly interacting probes placed at the
two end points. Such a model is described by the Hamiltonian
\begin{eqnarray}
\label{QLDEHamiltoniana}
H & = & J\sum_{i=2}^{N-2}\left[S_{i}^{x}S_{i+1}^{x}+S_{i}^{y}S_{i+1}^{y} \right] \\
& &  +  J\lambda \left(S_{1}^{x}S_{2}^{x}+S_{1}^{y}S_{2}^{y}+
S_{N-1}^{x}S_{N}^{x}+S_{N-1}^{y}S_{N}^{y} \right)  \; , \nonumber
\end{eqnarray}
where $0<\lambda<1$. The evolution of the end-to-end concurrence
shared by the end points as a function of the length of the chain
for different values of the weak coupling $\lambda$ is reported in
Fig.\ref{fig:njp5}.
At variance with what occurred in the models with patterns of
alternating couplings studied in the previous subsection, in the
case of models with uniform bulk interactions and weak end probes
the GS end-to-end entanglement is sensitive to the size of the
chain. It decreases slowly as the length of the chain grows and
vanishes at a critical value of the length, which depends on the
value of the weak coupling $\lambda$. This behavior of the GS
end-to-end entanglement is due to the fact that the region
responsible of preventing the formation of quantum correlations
between the end points and the bulk of the chain is fixed at two
lattice spacings, and does not grow as the length of the chain is
increased. However, by choosing sufficiently small values of the
weak coupling $\lambda$, it is always possible to realize a GS with
nonvanishing LDE in chains of arbitrary finite length. Hence it is
natural to name this behaviour as quasi LDE (QLDE), a phenomenon
that differs from perfect LDE, as the latter occurs only when the
system admits a non vanishing GS end-to-end entanglement that
attains its maximum (saturation) value in the thermodynamic limit.

In addition to the QLDE nature of the end-to-end entanglement in the
GS, models with weak end bonds possess an energy gap between the GS
and the first excited states that exhibits a very different behavior
compared to the case of models characterized by LDE.
\begin{figure}
\begin{centering} \includegraphics[width=75mm,keepaspectratio]{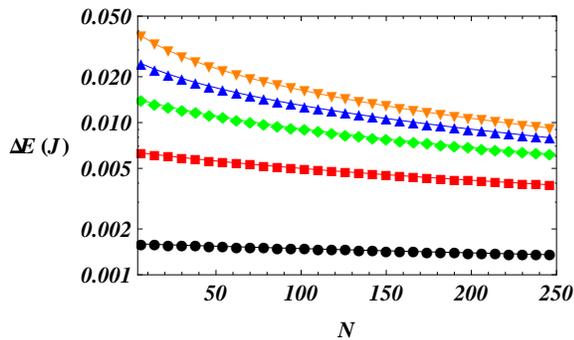}
\par\end{centering}
\caption{Energy Gap $\Delta E (J)$, in units of the coupling energy
$J$, between the GS and the first excited states for an $XX$ spin
chain with weak end probes, described by the Hamiltonian
Eq.(\ref{QLDEHamiltoniana}), plotted as a function of the length $N$
of the chain for different values of the weak coupling $\lambda$.
From top to bottom: orange inverted triangles: $\lambda=0.2$; blue
triangles: $\lambda=0.16$; green diamonds: $\lambda=0.12$; red full
squares: $\lambda=0.08$; black full circles: $\lambda=0.04$.}
\label{fig:njp6}
\end{figure}
As one can see from Fig. \ref{fig:njp6}, the energy gap does not
exhibit an exponential decay with the size of the chain as in the
cases of Fig. \ref{fig:njp2} and Fig. \ref{fig:njp4}. This fact,
together with the possibility to choose appropriately the value of
$\lambda$ to ensure a non vanishing QLDE shared by the end points in
chains of arbitrary finite length, opens the possibility for the
experimental realization of QLDE at low but realistically achievable
temperatures with concrete physical systems engineered in
configurations suitable for the realization/simulation of model
Eq.(\ref{QLDEHamiltoniana}).

\begin{figure}
\begin{centering} \includegraphics[width=75mm,keepaspectratio]{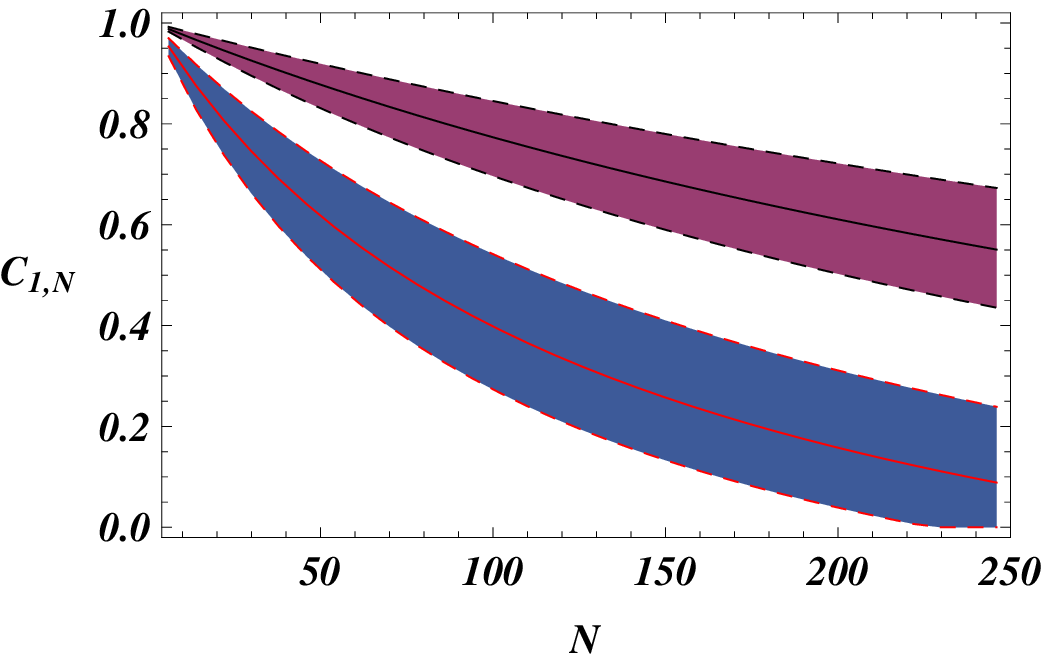}
\par\end{centering}
\caption{ End-to-end entanglement as a function of the length of the
chain for the model with disordered couplings described by \eq{QLDEHamiltoniana-noise}
where $\chi_i$ is a random variable uniformly distributed in the interval
$\{-\bar{\chi},\bar{\chi}\} $ with $\bar{\chi}=0.2$. The violet
area corresponding to $\lambda=0.2$, and the blue one corresponding
to $\lambda=0.4$, are the domains in which the different random samples may fall.
They have been obtained with $10^{4}$ independent samples for each value $N$
of the length of the chain.} \label{fig:njp5-err}
\end{figure}


Even if the models introduced in the previous subsection and the
one described by \eq{QLDEHamiltoniana} show different behaviors of
the end to end entanglement, they share the property that it is
robust in the presence of noise. Indeed, \fig{fig:njp5-err} displays
the end-to-end entanglement for different realizations of the model
with disordered couplings described by the following Hamiltonian:
\begin{eqnarray}
\label{QLDEHamiltoniana-noise}
H & = & J\sum_{i=2}^{N-2}\chi_i \left[S_{i}^{x}S_{i+1}^{x}+S_{i}^{y}S_{i+1}^{y} \right] \\
& &  +  J(\lambda
+\chi_1)\left(S_{1}^{x}S_{2}^{x}+S_{1}^{y}S_{2}^{y}
\right)+ \nonumber \\
 & & J(\lambda +\chi_{N-1}) \left(S_{N-1}^{x}S_{N}^{x}+S_{N-1}^{y}S_{N}^{y} \right)  \; ,
\nonumber
\end{eqnarray}
where $\chi_i$ is a random variable uniformly distributed in the interval
$\{-\bar{\chi},\bar{\chi}\} $. As with models endowed with perfect
LDE, also in the case of QLDE the presence of noise affects the end to end
concurrence bi-directionally, and the entanglement remains nonvanishing
even when the relative weight of the noise is up to $20 \%$.
Also in the case of models associated to QLDE, about half of the
random sample realizations happen to increase the value of the
concurrence between the end points. Moreover, the disruptive
effect of noise is strongly reduced for large values of the
end-to-end entanglement.

\begin{figure}
\begin{centering} \includegraphics[width=75mm,keepaspectratio]{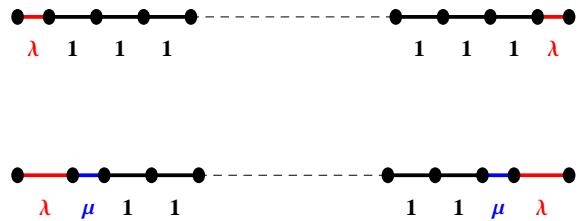}
\par\end{centering}
\caption{Schematic drawing of $XX$ spin chains endowed with QLDE in
the GS. Upper drawing: Model Eq.(\ref{QLDEHamiltoniana}), with weak
end bonds $\lambda <1$ and uniform unit couplings in the bulk. Lower
drawing: $\lambda$--$\mu$ model Eq.(\ref{QLDE-1-Hamiltoniana}), with
weak end bonds $\lambda <1$, strong near-end bonds $\mu >1$, and
uniform unit couplings in the bulk.} \label{fig:ModelliQLDE}
\end{figure}
\begin{figure}
\begin{centering} \includegraphics[width=75mm,keepaspectratio]{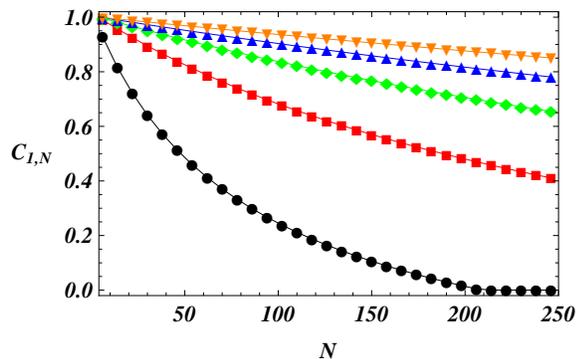}
\par\end{centering}
\caption{End-to-end concurrence $C_{1,N}$ for the $\lambda$--$\mu$
spin chain described by Hamiltonian Eq.(\ref{QLDE-1-Hamiltoniana}),
plotted as a function of the length $N$ of the chain, for a fixed
value of the weak coupling $\lambda = 0.1$, and different values of
the strong coupling $\mu$. From top down: orange inverted triangles:
$\mu=5$; blue triangles: $\mu=4$; green diamonds: $\mu=3$; red full
squares $\mu=2$. The lowest-lying line (black full circles) is the
one with $\mu=1$, corresponding to the $\lambda$ model
Eq.(\ref{QLDEHamiltoniana}), and is drawn for comparison.}
\label{fig:njp7}
\end{figure}
\begin{figure}
\begin{centering} \includegraphics[width=75mm,keepaspectratio]{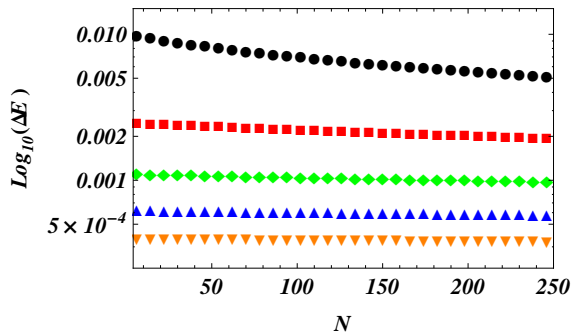}
\par\end{centering}
\caption{Energy Gap $\Delta E (J)$ (in logarithmic scale), in units
of the uniform bulk coupling $J$, between the GS and the first
exited states for the $\lambda$--$\mu$ spin chain described by
Hamiltonian Eq.(\ref{QLDE-1-Hamiltoniana}), plotted as a function of
the length $N$ of the chain, for a fixed value of the weak coupling
$\lambda = 0.1$, and different values of the strong coupling $\mu$.
From bottom up: orange inverted triangles: $\mu=5$; blue triangles:
$\mu=4$; green diamonds: $\mu=3$; red full squares $\mu=2$. The line
with $\mu=1$ (black full circles) corresponds to the $\lambda$ model
Eq.(\ref{QLDEHamiltoniana}), and is drawn for comparison.}
\label{fig:njp8}
\end{figure}

The role played by the two weak interactions in the model
Eq.(\ref{QLDEHamiltoniana}) is to force the second ($N-1$-th) spin
of the chain to get heavily entangled with the third ($N-2$-th)
spin, thus forcing the first and last spins of the chain, due to
entanglement monogamy and the fact that this configuration is
energetically favorable, to develop a nonvanishing quantum
correlation in the GS. This effect can be enhanced further by
increasing the interaction between the second and the third spin,
and correspondingly between the $N-1$-th and the $N-2$-th, well
above the reference value between neighboring spins in the bulk of
the chain. In this case the model Hamiltonian reads:
\begin{eqnarray}
\label{QLDE-1-Hamiltoniana}
H & = & J\sum_{i=3}^{N-3}\left[S_{i}^{x}S_{i+1}^{x}+S_{i}^{y}S_{i+1}^{y} \right] \\
& &  +  \lambda \left(S_{1}^{x}S_{2}^{x}+S_{1}^{y}S_{2}^{y}+
S_{N-1}^{x}S_{N}^{x}+S_{N-1}^{y}S_{N}^{y} \right) \; , \nonumber
\\ & &  + \mu \left(S_{2}^{x}S_{3}^{x}+S_{2}^{y}S_{3}^{y}+
S_{N-2}^{x}S_{N-1}^{x}+S_{N-2}^{y}S_{N-1}^{y} \right)  \; .
\nonumber
\end{eqnarray}
Models Eq.(\ref{QLDEHamiltoniana}) and
Eq.(\ref{QLDE-1-Hamiltoniana}) with their spatial patterns of
site-dependent couplings are sketched pictorially in Fig.
\ref{fig:ModelliQLDE}.
In Eq.(\ref{QLDE-1-Hamiltoniana}) we have $\lambda < 1$ and $\mu >
1$, so that $\lambda J$ (weak end bond) $< J$ (uniform bulk
interaction) $< \mu J$ (strong near-end bond). The increased
efficiency granted by model Eq.(\ref{QLDE-1-Hamiltoniana}) for the
process of creating a nonvanishing QLDE between the two end points
of the chain can be appreciated in Fig. \ref{fig:njp7}, where the
model (\ref{QLDEHamiltoniana}) with simple weak end bonds $\lambda =
0.1$, $\mu = 1$, is compared with model (\ref{QLDE-1-Hamiltoniana})
with the same value of $\lambda$ and different values of the strong
near end bonds $\mu > 1$. In the following, Hamiltonians
(\ref{QLDEHamiltoniana}) realizing simple QLDE in the GS will be
referred to as $\lambda$ models, while Hamiltonians
(\ref{QLDE-1-Hamiltoniana}) that realize enhanced QLDE in the GS
will be referred to as $\lambda$--$\mu$ models.
It is not surprising that the strong enhancement brought by the
$\lambda$--$\mu$ model to the QLDE that can be accommodated in the
GS is obtained at the cost of a (relatively moderate) trade-off with
the behaviour of the energy gap. In Fig. \ref{fig:njp8} we have
compared the behaviour of the energy gap, as a function of the size
of the system, for the $\lambda$ model and various $\lambda$--$\mu$
models with the same value of $\lambda$ and different values of
$\mu$.
At this point, the question naturally arises of optimizing the
parameters of the Hamiltonian in order to single out the maximum
possible value of the QLDE compatible with an energy gap
sufficiently large to warrant a concrete physical realizability at
finite temperature. Obviously, the process of optimization of the
Hamiltonian parameters does not provide a unique general answer.
Rather, the results will vary, depending on the length of the chain
that one wants/needs to consider as well as on the minimum working
temperature that is fixed by the external experimental conditions.
In the following we will always consider the models of enhanced QLDE
described by Hamiltonian (\ref{QLDE-1-Hamiltoniana}), that for
$\mu=1 $ reduce to the models (\ref{QLDEHamiltoniana}) of simple
QLDE.
\begin{figure}
\begin{centering} \includegraphics[width=80mm,keepaspectratio]{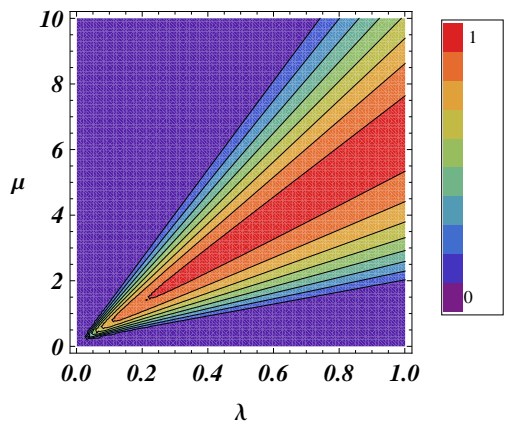}
\par\end{centering}
\caption{Two-dimensional contour plot of the end-to-end concurrence
$C_{1,N}$ as a function of the weak and strong couplings $\lambda$
and $\mu$, for a 1-D $\lambda$--$\mu$ spin model
(\ref{QLDE-1-Hamiltoniana}) on an open linear chain of $N=10$ spins
at a reduced temperature $T/J=0.05$. The color map relative to the
values of the concurrence is reported on the left. Values vary from
$0$ (violet) to $1$ (red).} \label{fig:njp9}
\end{figure}
A typical result of the optimization process for a $\lambda$--$\mu$
spin chain of finite length is shown in Fig. \ref{fig:njp9} in the
case $N = 10$. It is important to notice that too high values of
$\mu$ and/or too low values of $\lambda $, contrary to naive
intuition do not help, because they lead either to an effective
separation of the end points from the rest of the chain and/or to a
much too small energy gap compared to the energy amplitude of
thermal excitations. In both cases, the associated QLDE vanishes.
Moreover, the fact that very large values of the end-to-end
entanglement are always associated with small energy gaps implies,
as the figure shows, that the optimization process is never trivial
(especially in determining the regions of matching values for the
couplings), and therefore it represents a key point in the
discussion of realistic physical systems able to simulate/realize
models endowed with QLDE, a subject that we investigate in the
following sections.

\begin{figure}
\begin{centering} \includegraphics[width=75mm,keepaspectratio]{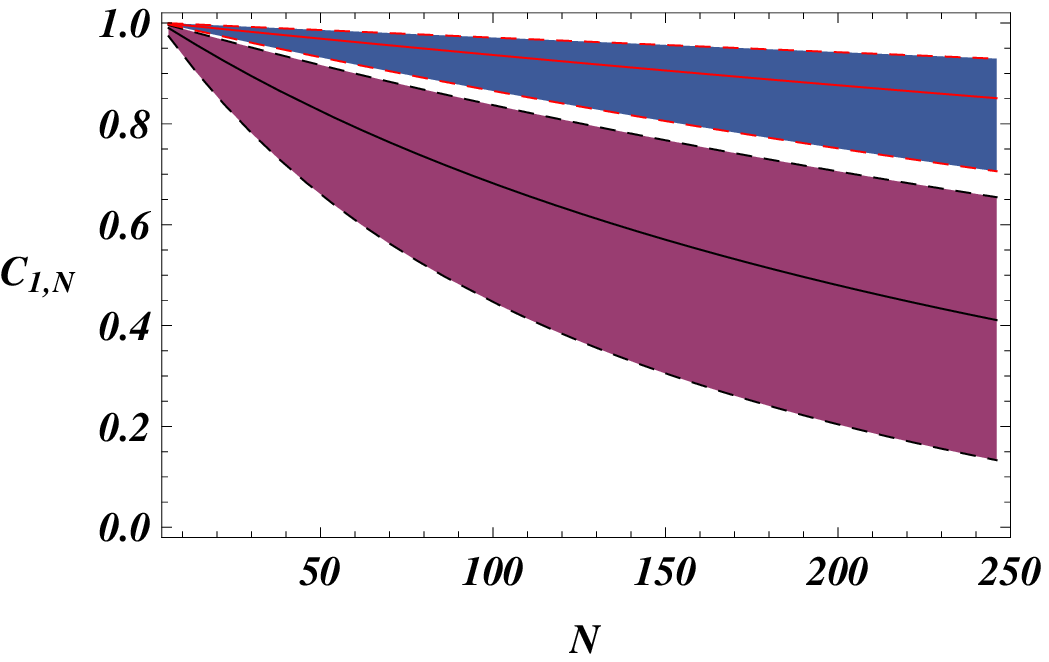}
\par\end{centering}
\caption{ End-to-end entanglement as a function of the length of the
chain for the model with disordered couplings described by \eq{QLDE-1-Hamiltoniana-noise}
where $\chi_i$ is a random variable uniformly distributed in the interval
$\{-\bar{\chi},\bar{\chi}\} $ with $\bar{\chi}=0.2$. The
violet area, corresponding to $\lambda=0.1$ and $\mu=2.0$ and
the blue area, corresponding to $\lambda=0.1$ and $\mu=5.0$,
are the domains in which the different random samples may fall.
They have been obtained with $10^{4}$ independent samples for
each value of the length $N$ of the chain.}
\label{fig:njp7-err}
\end{figure}


As with the other cases considered previously, also the $\lambda-\mu$ model is robust against
noise affecting the coupling amplitudes. This resilience is shown in
\fig{fig:njp7-err}, displaying the end-to-end entanglement
for different realizations of the model with disordered couplings
described by the Hamiltonian:
\begin{eqnarray}
\label{QLDE-1-Hamiltoniana-noise}
H & = & J\sum_{i=3}^{N-3} \chi_i \left[  S_{i}^{x}S_{i+1}^{x}+S_{i}^{y}S_{i+1}^{y} \right] \\
& &  + \lambda
\chi_1\left(S_{1}^{x}S_{2}^{x}+S_{1}^{y}S_{2}^{y}\right)
 \nonumber \\
& & +\lambda \chi_{N-1}\left(
S_{N-1}^{x}S_{N}^{x}+S_{N-1}^{y}S_{N}^{y}\right)  \; ,
\nonumber \\
& &  + \mu \chi_{2}
\left(S_{2}^{x}S_{3}^{x}+S_{2}^{y}S_{3}^{y}\right) \nonumber \\
& & +\mu \chi_{N-2} \left(
S_{N-2}^{x}S_{N-1}^{x}+S_{N-2}^{y}S_{N-1}^{y} \right)  \; .
\nonumber
\end{eqnarray}
Here, as in the previous cases, $\chi_i$ is a random variable
uniformly distributed in the interval $\{-\bar{\chi},\bar{\chi}\}$.


\section{End-to-end entanglement in optical lattices}

Until now we have analyzed in some detail the question of
identifying specific instances of open $XX$ spin chains with
different patterns of site-dependent interactions that possess GSs
endowed with large values of the entanglement shared by the end
points and an energy gap compatible with physical realizations at
finite temperature. Here and in the following we investigate
concrete instances of atomic and quantum optical systems that may
allow the experimental demonstration of LDE and QLDE.

We first consider systems of ultracold neutral atoms loaded on a
one-dimensional optical lattice generated by interfering laser beams
\cite{Bloch}. The dynamics of such systems is usually well
described, depending on the nature of the atoms present in the
ensemble, by a Bose-Hubbard model \cite{Jaksch} or by a Fermionic
Hubbard model, or by mixtures of atoms of different species, such as
Bose-Bose \cite{Buonsante}, Fermi-Fermi \cite{Iskin}, and Bose-Fermi
mixtures \cite{Illuminati}. Regardless of the specific cases
realized, the parameters in the various Hamiltonians can be easily
controlled by tuning both the intensity and the frequency of the
laser beams. This fact, i.e. that all the Hamiltonian parameters can
be manipulated and modified with a high degree of control, makes
optical lattices of particular interest for the
simulation/realization of models of interacting quantum systems. The
mapping of Hubbard-type models into $XX$ spin models, in appropriate
regimes of the parameters, was originally discussed in the
fundamental work of Fisher {\em et al.} \cite{Fisher} where the
Bose-Hubbard model was introduced. In recent years, several works
have been dedicated to the simulation of spin models using Hubbard
Hamiltonians realized in optical lattices. However, in most of these
works the spin-spin interactions are usually taken to be either
uniform \cite{Duan,Kuklov} or completely random \cite{Sanpera}.

Let us consider a 1-D optical chain loaded with single-species
ultracold bosonic atoms. The system is described by the 1-D
Bose-Hubbard Hamiltonian \cite{Jaksch} that can be written as
\begin{equation}\label{BHHamiltoniana}
 H= \frac{U}{2}\sum_{i=1}^N n_i(n_i-1) -t\sum_{i=1}^{N-1} (b_i^\dagger
 b_{i+1}+h.c.) \; ,
\end{equation}
where $b_i^\dagger$ ($b_i$) is the creation (annihilation) operator
of a bosonic atom at the $i$-th site of the chain, $n_i=b_i^\dagger
b_i$ is the number operator at site $i$, and $N$ is the number of
sites in the chain. The parameter $U$ denotes the strength of the
local on-site repulsion, while $t$ is the hopping amplitude between
adjacent sites. These Hamiltonian parameters depend on the
quantities characterizing the external periodic optical field,
taking into account only the lowest vibrational states for every
minimum of the periodic potential \cite{Jaksch}.

The simplest approach to simulate $XX$ spin models starting with the
bosonic Hubbard Hamiltonian Eq.(\ref{BHHamiltoniana}) relies on the
fact that the local Fock states with either one or no particle per
site differ from all other local Fock states in that they are the
only two local levels whose energy is not dependent on the
repulsion. Therefore, in the strong interaction regime $(U \gg t)$,
they are automatically separated from the others, and hence they can
play the role local spin-$1/2$ states. In the framework of this
approximation, the hopping term is naturally mapped in an
interaction of the $XX$ type with an amplitude equal to two times
the hopping amplitude. However, in an optical lattice, the hopping
amplitude between adjacent sites, depends both on the frequency and
the intensity of the lasers that are quantities common to each site.
Hence, the interaction couplings between adjacent spins in the
ensuing $XX$ model are site-independent.

\begin{figure}[t]
\includegraphics[width=6.5cm]{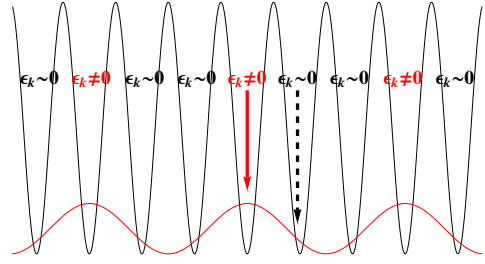}
\caption{Scheme of the optical super-lattice potential obtained by
combining two appropriate sets of stationary laser beams. It
realizes an effective $XX$ spin chain with alternating weak and
strong interactions that allows for LDE between the two end points.
The first set of laser beams generates an optical potential (black
line) that realize an optical lattice. On the contrary, the weaker
optical potential of the second sets, represented by the red line,
creates a local field at each site of the optical lattice. The local
field is strongly enhanced on every site over
three.}\label{modulazione}
\end{figure}

\begin{figure}[t]
\includegraphics[width=7.5cm]{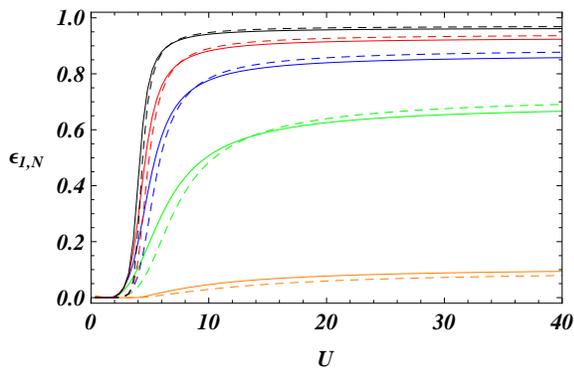}
\caption{End-to-end entanglement in the GS of a super-lattice system
of $N=12$ sites and $n=4$ atoms, with strong local field
$\varepsilon$ on $2-nd$, $5-th$, $8-th$, and $11-th$ site, plotted
as a function of the on-site repulsion $U$ in units of the hopping
amplitude $t$. The dashed lines represent the logarithmic
negativity, measuring the end-to-end entanglement in the original
Bose-Hubbard atomic system, while the solid lines stand for the
concurrence, that measures the end-to-end entanglement in the
effective $XX$ spin model with alternating weak and strong
couplings. The different curves correspond to different values of
the strong local field $\varepsilon$. From bottom up: orange:
$\varepsilon = t$; green: $\varepsilon = 3t$; blue: $\varepsilon =
5t$; red: $\varepsilon = 7t$; black: $\varepsilon = 10t$.}
\label{njp-b3}
\end{figure}

However, site-dependent spin-spin couplings can be easily engineered
by introducing local fields on properly selected sites of the
lattice. Consider a local field on a given site, say the $k$-th one,
whose amplitude is larger than the hopping amplitudes but weaker
than the on-site repulsion $U$. In this situation occupation of the
on-site single-particle Fock state is energetically unfavorable.
Therefore, if an atom in $k+1$-th or in $k-1$-th site hops onto the
$k$-th site, it is immediately pushed away. This mechanism realizes
either a self-interaction at site $k$ or an effective interaction
between the $k+1$-th and the $k-1$-th site. This simple reasoning
can be put in a quantitative form resorting to degenerate
perturbation theory in powers of the hopping amplitude. Let $k-1$,
$k$, and $k+1$ be three adjacent sites, let $\varepsilon_k\neq0$ the
local field at site $k$, and such that $t\ll \varepsilon_k \ll U$.
In second-order perturbation, discarding all the excited states with
energy proportional to $U$ and $\varepsilon_k$, each pair term in
Eq.(\ref{BHHamiltoniana}) maps into the following spin-spin
interaction:
\begin{eqnarray}\label{sitedependence}
    H_k&=&-\frac{2
    t^2}{\varepsilon_k}(S^x_{k-1}S^x_{k+1}+S^y_{k-1}S^y_{k+1})
    \nonumber \\
    & &- \frac{t^2}{\varepsilon_k}\left(S^z_{k-1}+\frac{1}{2}\right)^2
    - \frac{t^2}{\varepsilon_k}\left(S^z_{k+1}+\frac{1}{2}\right)^2 \; .
\end{eqnarray}
It is important to note that in these spin-spin interaction terms
the dependence on the $k$-th site is removed. Hence, the
introduction of site-dependent spin-spin interaction strengths is
associated to a reduction of the sites of the optical lattice.
Eq.(\ref{sitedependence}) is the basic ingredient in the realization
of $XX$ spin chains with site-dependent interactions and
nonvanishing end-to-end entanglement. The successive task is to
determine suitably engineered optical lattices with the appropriate
local field dynamics. For instance, one possible way to realize the
$XX$ spin model with alternating weak and strong couplings
Eq.\ref{LDEHamiltonian} is to introduce an optical super-lattice
potential obtained through the interference of two sets of laser
beams of different amplitude strengths. The set of stronger beams
realizes the lattice structure while the set of weaker ones realizes
and modulates the presence of a local field at each lattice site.
If, for instance, the set of weak beams is tuned at a wavelength
three times as large as that of the set of strong ones, one realizes
a system in which every site over three is characterized by a strong
local field, while the remaining two sites experience a practically
vanishing local field, as schematically illustrate in Fig.
\ref{modulazione}.
Resorting to degenerate second-order perturbation theory in powers
of $t$, and taking into account Eq.(\ref{sitedependence}), the
optical lattices Hamiltonian is mapped in the spin Hamiltonian
\begin{eqnarray}
H &=& - t \sum_{i=1}^{N'-1}
\left[\left(1+\frac{t}{\varepsilon}\right)+(-1)^i\left(1-\frac{t}{\varepsilon}\right)\right]
\times \; \nonumber \\
& & \left(S_{i}^{x}S_{i+1}^{x} + S_{i}^{y}S_{i+1}^{y}\right) -
\frac{t^2}{\varepsilon} \sum_{i=1}^{N'}
\left(S^z_{i}+\frac{1}{2}\right)^2 ,
 \label{LDEHamiltonianopticallattices}
\end{eqnarray}
where $\varepsilon > t$ is the maximum amplitude of the local field,
and $N'=2N/3$ is the number of spins in the chain that, as already
noticed, does not coincide with the number of sites in the physical
optical lattice chain. The last sum of terms represents an overall
energy offset that can always be reabsorbed in the definition of the
zero of the energy and has no dynamical effect. Fixing the total
magnetization at $0$, corresponds to fixing the number of atoms in
the optical lattice at $n=N'/2=N/3$. In this case we obtain that in
the strong interaction regime the optical super-lattice simulates
the 1-D $XX$ spin chain with alternating weak and strong couplings
Eq.\ref{LDEHamiltonian}, exactly with $\lambda = t/\varepsilon$. It
is important to verify the soundness of the theoretical mapping,
obtained in second-order perturbation theory, by direct exact
numerical comparison between the end-to-end entanglement properties
of the spin model Eq.\ref{LDEHamiltonian} and of the original
bosonic system. In Fig. \ref{njp-b3} we report the end-to-end
entanglement in the GS of a system of ultracold bosonic atoms loaded
on the above-described optical super-lattice, with $N=12$ sites and
$n=N/3=4$ atoms, corresponding, according to the our previous
analysis, to a 1-D $XX$ spin chain with $N'=2N/3=8$ sites with
alternating weak and strong couplings. As measures of entanglement,
we consider the concurrence \cite{Wootters} and the logarithmic
negativity \cite{Logneg}. The concurrence quantifies the GS
entanglement between the two end points seen as effective spins,
obtained by tracing out all Fock states but the empty state and the
state with single-atom occupation. The logarithmic negativity is
used to measure the actual GS entanglement of the full bosonic
system between the two end sites of the optical super-lattice.
Fig. \ref{njp-b3} shows that for small values of the on-site
repulsion $U$ there is no end-to-end entanglement regardless of the
value of the local field $\varepsilon$. Viceversa, in the strong
interaction regime the system develops a sizeable end-to-end
entanglement that always approximates very closely the end-to-end
entanglement of the $XX$ spin chain with alternating weak and strong
couplings, proving that appropriate optical lattice or super-lattice
structures can simulate/realize efficiently $XX$ spin chains with
LDE entanglement. An exception is constituted by the limiting case
$\varepsilon=t$. In this situation the strong local field condition
$\varepsilon > t$ is violated, $\lambda=1$ in the corresponding spin
Hamiltonian, and thus the end-to-end spin-spin concurrence vanishes,
while the end-to-end logarithmic negativity in the GS of the
original optical super-lattice system is definitely nonvanishing.

We now turn to the question of simulating spin models with QLDE
rather than pure LDE. One problem in simulating the Hamiltonians
described in Subsection \ref{QLDE} with atomic systems in optical
potential structures is that QLDE models do not have a periodic
pattern of site-dependent interactions and thus are difficult to
approximate with super-lattices, as done for LDE models. Spin models
with QLDE structures can be simulated by addressing the degrees of
freedom of the individual lattice sites with pre-selection
techniques that have recently been discussed for different purposes
\cite{Das Sarma}, \cite{Cho}. If these methods, or similar ones,
could be implemented successfully in realistic experimental
conditions, it would be possible to realize optical-lattice
structures such that only few pre-selected sites are affected by a
nonvanishing local field. Consider the following Bose-Hubbard
Hamiltonian
\begin{eqnarray}\label{BHHamiltoniana-1}
 H &= &\frac{U}{2}\sum_{i=1}^N n_i(n_i-1) -t\sum_{i=1}^{N-1} (b_i^\dagger
 b_{i+1}+h.c.) \nonumber \\ & & + \varepsilon (n_2+n_{N-1})\; .
\end{eqnarray}
The presence of a local field $\varepsilon$ at the second and
$N-1$-th sites allows to simulate, when $ t < \varepsilon < U$, the
$XX$ spin model with weak end probes described by
Eq.(\ref{QLDEHamiltoniana}).
\begin{figure}[t]
\includegraphics[width=7.cm]{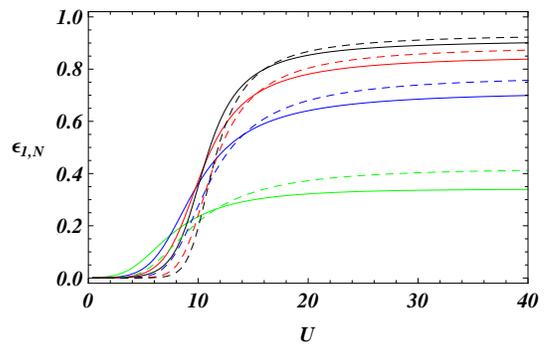}
\caption{End-to-end entanglement in the GS of the Bose-Hubbard
Hamiltonian Eq. (\ref{BHHamiltoniana-1}), defined on an optical
lattice of $N = 10$ sites with strong local field on the $2-nd$ and
on the $9-th$ site, plotted as a function of the on-site repulsion
$U$ in units of the hopping amplitude $t$. The dashed lines
correspond to the logarithmic negativity measuring the end-to-end
entanglement in the original Bose-Hubbard atomic system, while the
solid lines stand for the concurrence, that measures the end-to-end
entanglement in the effective $XX$ spin model with weak end probes.
The different curves correspond to different values of the local
field $\varepsilon$. From bottom up: green: $\varepsilon = 3t$;
blue: $\varepsilon = 6t$; red: $\varepsilon = 9t$; black:
$\varepsilon = 12t$.} \label{njp-b1}
\end{figure}
In Fig.(\ref{njp-b1}) we report the GS end-to-end entanglement for
the Hamiltonian Eq. (\ref{BHHamiltoniana-1}) defined on an optical
lattice of $N=10$ sites and loaded with $n=4$ bosonic atoms, and
compare it with the GS end-to-end entanglement of an $XX$ model with
weak end probes Eq.(\ref{QLDEHamiltoniana}) defined on a chain with
$8$ spins. We see that the former is an excellent approximation to
the latter in the limit of large on-site repulsion $U$. We have
fixed the total number of bosonic atoms to (N-2)/2 to satisfy the
half-filling condition. This corresponds again to impose a vanishing
magnetization in the effective spin chain. In the strong-coupling
regime, the value of the end-to-end entanglement grows as the local
field $\varepsilon$ is increased, as should be expected from the
results of the discussion in Subsection \ref{QLDE}, taking into
account Eq. (\ref{sitedependence}).
\begin{figure}[t]
\includegraphics[width=7.cm]{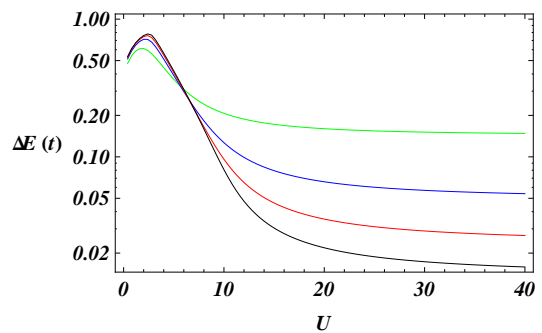}
\caption{Energy gap between the GS and the first excited states of
the Bose-Hubbard Hamiltonian Eq. (\ref{BHHamiltoniana-1}), defined
on an optical lattice of $N = 10$ sites with strong local field on
the $2-nd$ and on the $9-th$ site, plotted as a function of the
on-site repulsion $U$ in units of the hopping amplitude $t$. The
different curves correspond to different values of the local field
$\varepsilon$. From bottom up: green: $\varepsilon = 3t$; blue:
$\varepsilon = 6t$; red: $\varepsilon = 9t$; black: $\varepsilon =
12t$.} \label{njp-b2}
\end{figure}
However, as discussed in the previous section, the significance of
simulating/realizing spin models with QLDE using, for instance,
optical lattice Hamiltonian structures as the ones described by Eq.
(\ref{BHHamiltoniana-1}) is that the energy gap can be sufficiently
large to ensure that a nonvanishing end-to-end entanglement is still
retained at very low but realistically reachable working
temperatures. The two requirements of large values of QLDE and a
large energy gap are partly conflicting. The behavior of the energy
gap is reported in Fig.\ref{njp-b2}. Comparing Figs.\ref{njp-b1} and
\ref{njp-b2} we see that the optimal compromise, that allows to
obtain both a sizeable QLDE and a sizeable energy gap, takes place
in initial side of the strong-coupling regime, at moderately large
values of the on-site repulsion $U$.

\section{End-to-end entanglement in arrays of coupled optical cavities}

Recently, hybrid atom-optical systems of coupled cavity arrays
(CCAs) have been intensively studied in relation to their ability to
simulate collective phenomena typical of strongly correlated systems
\cite{Hartmann,Greentree,Angelakis,Rossini,Hartmann-2}. In the
present section we will discuss in detail some recent findings
\cite{Salerno2} about the possibility of exploiting appropriately
engineered arrays of coupled optical cavities to realize $XX$ open
spin chains sustaining LDE or QLDE. Besides the extremely high
controllability and the straightforward addressability of single
constituents, arrays of coupled cavities also allow in principle a
great degree of flexibility in their design and geometry
\cite{Plenio1}. Therefore, they should be tested as natural
candidates for the realization of spatially extended communication
networks and scalable computation devices.

\begin{figure}[t]
\includegraphics[width=6.cm]{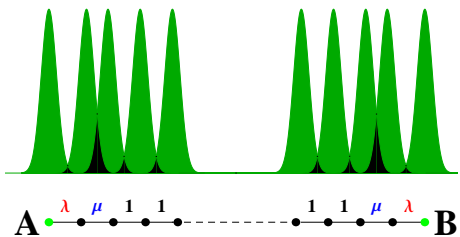}
\caption{Scheme of an array of coupled optical cavities realizing a
$\lambda$--$\mu$ linear spin chain described in Eq.
(\ref{QLDE-1-Hamiltoniana}). Dark green: Area covered by the wave
functions associated to each site of the array. The two next-to-end
sites (light green circles) are symmetrically displaced with respect
to a situation of perfectly equispaced sites, and drawn closer to
their neighbors in the bulk (black circles). As a consequence, the
overlap (black area) between the wave functions of these two
cavities and their neighbors in the bulk is larger than the would-be
reference (unit) overlap in an equispaced array. At the same time,
the overlap between the end sites of the array (red and blue
circles) and the next-to-end sites is reduced proportionally
compared to an equispaced array.} \label{CCA}
\end{figure}

Consider a linear CCA with open ends, consisting of $N$ cavities.
The dynamics of a single constituent of the array doped with a
single two-level atom is well described by the Jaynes-Cummings
Hamiltonian \cite{Jaynes}
\begin{equation}
\label{Hamiltoniana locale}
H_k = \omega a^\dagger_k a_k+\omega' \ket{e_k}\bra{e_k} +
g  a^\dagger_k \ket{g_k}\bra{e_k} + g\ket{e_k}\bra{a_k} a_k \, ,
\end{equation}
where $a_k$ ($a^\dagger_k$) is the annihilation (creation) operator
of photons with energy $\omega$ in the $k$-th cavity, $\ket{g_k}$
and $\ket{e_k}$ are respectively the ground and excited atomic
states, separated by the gap $\omega'$, and $g$ is the photon-atom
coupling strength. The local Hamiltonian \eq{Hamiltoniana locale} is
immediately diagonalized in the basis of dressed photonic and atomic
excitations (polaritons):
\begin{eqnarray}\label{autostati}
 \ket{\emptyset_k} & = & \ket{g_k}\ket {0_k} \, ;\\
 \ket{n+_k} & = & \cos\theta_n \ket{g_k} \ket{n_k} + \sin \theta_n
 \ket{e_k}\ket{(n-1)_k} \; \; n \ge 1\,  ; \nonumber \\
 \ket{n-_k} & = & \sin\theta_n \ket{g_k}\ket{n_k} - \cos \theta_n
 \ket{e_k}\ket{(n-1)_k} \; \; n \ge 1\, , \nonumber
\end{eqnarray}
where $\theta_n$ is given by $\tan 2 \theta_n=-g\sqrt{n}/\Delta$ and
$\Delta=\omega'-\omega$ is the atom-light detuning. Each polariton
is characterized by an energy equal to
\begin{figure}[t]
\includegraphics[width=6.5 cm]{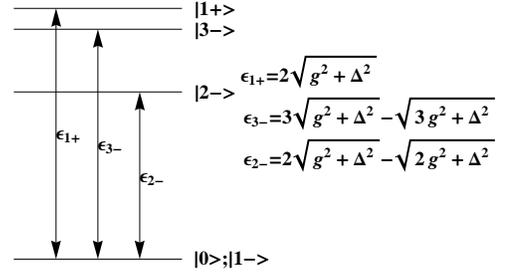}
\caption{Representation of the energy levels for a cavity with
$\omega = \sqrt{ g^2+\Delta^2}$. The GS is two-fold
degenerate, and the energy gap $\varepsilon_{2-}$ prevents the
occupancy of the higher energy levels, thus realizing an effective
two-level system.} \label{livelli}
\end{figure}
\begin{equation}\label{autoenergie}
 \varepsilon_0 = 0 ; \; \; \; \; \; \;
 \varepsilon_{n\pm} = n \omega \pm \sqrt{n g^2+\Delta^2} \,.
\end{equation}
When $\omega = \sqrt{ g^2+\Delta^2}$ the GS of \eq{Hamiltoniana
locale} becomes two-fold degenerate, resulting in a superposition of
$\ket{\emptyset_k}$ and $\ket{1-_k}$, see \fig{livelli}. If both the
atom-cavity interaction energy and the working temperature are small
compared to $\varepsilon_{2-}=2 \sqrt{ g^2+\Delta^2} - \sqrt{2
g^2+\Delta^2}$, one may neglect all the local polaritonic states but
$\ket{\emptyset_k}$ and $\ket{1-_k}$. This situation defines a local
two-level system. Adjacent cavities can be easily coupled either by
photon hopping or via wave guides of different dielectric and
conducting properties. The wave function overlap of two adjacent
cavities introduces the associated tunneling elements, so that the
total Hamiltonian of the CCA reads:
\begin{equation}
\label{Hamiltoniana totale} H_{cca} = \sum_k^N H_k  - \sum_k^{N-1}
J_{k} (a^\dagger_k a_{k+1} + a^\dagger_{k+1} a_{k}) \; .
\end{equation}
Each hopping amplitude $J_k$ depends strongly on both the geometry
of the cavities and the actual overlap between adjacent cavities. If
the maximum value among all the couplings $\{ J_k \}$ is much
smaller than the energy of the first excited state: $\max\{J_k\}\ll
\varepsilon_{2-}$, then the total Hamiltonian \eq{Hamiltoniana
totale} can be mapped in a spin-1/2 model of the $XX$ type with
site-dependent couplings of the form Eq.
(\ref{QLDE-1-Hamiltoniana}), where the state $\ket{\emptyset_k}$
($\ket{1-_k}$) plays the role of $\ket{\downarrow_k}$
($\ket{\uparrow_k}$). The mapping to an open-end $\lambda$--$\mu$
linear spin chain sustaining QLDE in the GS is then realized, e.g.,
by simply tuning the distance between the end- and next-to-end sites
of the array, as showed in \fig{CCA}.
The $\lambda$--$\mu$ $XX$ spin chain thus can be realized starting
from an equispaced CCA with site-independent nearest-neighbor bulk
coupling amplitude $J_b$, and then engineering appropriately the
positions of the $2-nd$ and of the $(N-1)-th$ cavities, that are
placed closer to their neighbors in the bulk and farther away from
the end points of the array. This shift lowers below unity the
reduced coupling between the end points of the array and their
next-to-end neighbors ($J_1/J_b=J_{N-1}/J_b=\lambda < 1$), and
increases above unity the reduced coupling between the next-to-end
sites and their neighbors in the bulk ($J_2/J_b=J_{N-2}/J_b=\mu >
1$), effectively realizing the $\lambda$--$\mu$ model of QLDE.
Obviously, adjusting the hopping rate by spacing the cavities
closer to the end points is not the only way to realize the
$\lambda-\mu$ model in an array of optical cavities. Indeed, it
can be technically rather challenging. An alternative way to
tune the inter-cavity hopping amplitudes, probably much more feasible
with currently available technologies, is to dope the cavities
with few-level atoms and use Raman transitions and the atomic
Lambda level structures to tune the couplings between two
cavities {\it in situ}, by using external laser drives as described,
e.g., in Ref. \cite{Hartmann-2}.

In the next section we will discuss the properties of QLDE, realized
using CCAs, in connection with the implementation of tasks of
quantum information science. In particular, we will discuss how the
$\lambda$--$\mu$ model and the associated GS QLDE can be exploited
to realize high-fidelity, long-distance quantum teleportation
protocols.

\section{Applications: QLDE and long-distance quantum teleportation in CCAs}

\begin{figure}[t]
{\includegraphics[width=3.1cm]{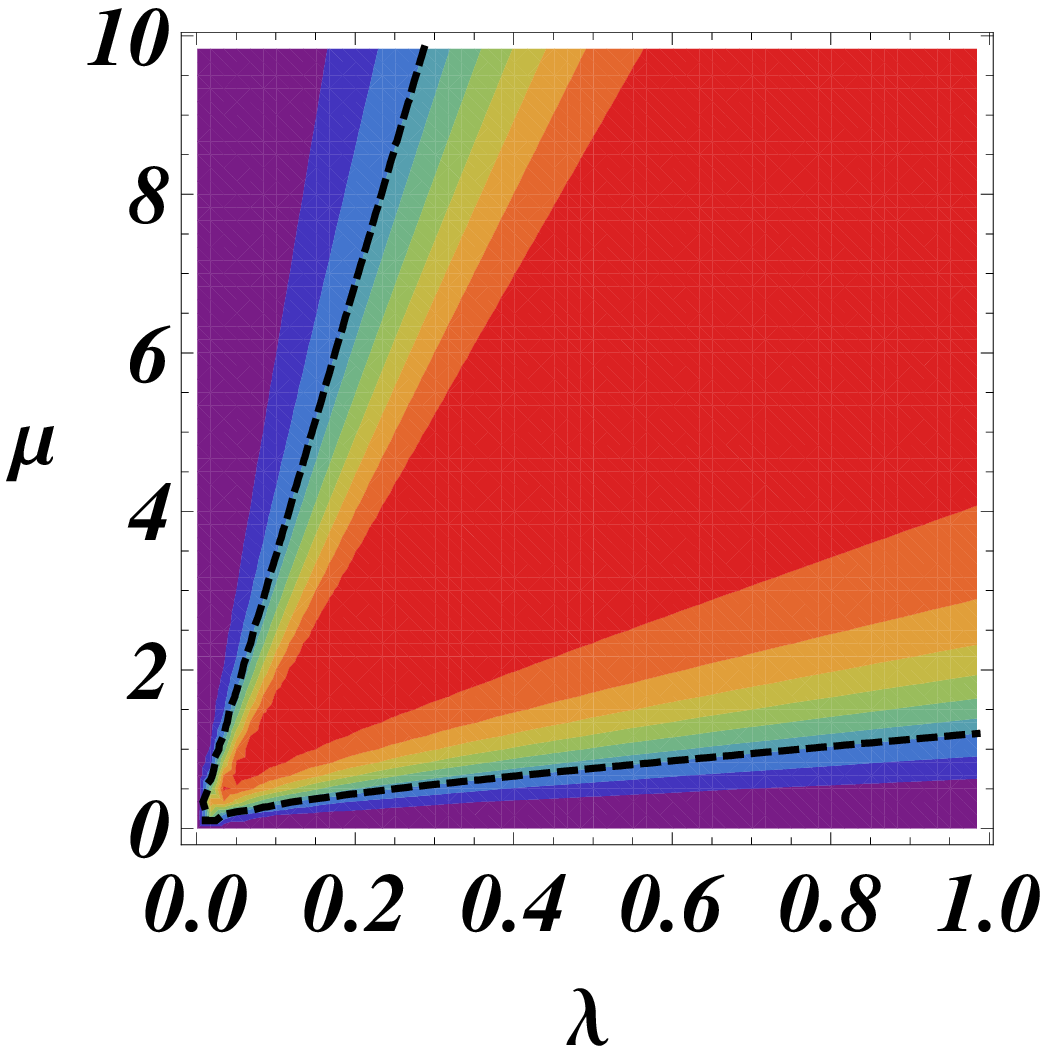}
\includegraphics[width=3.1cm]{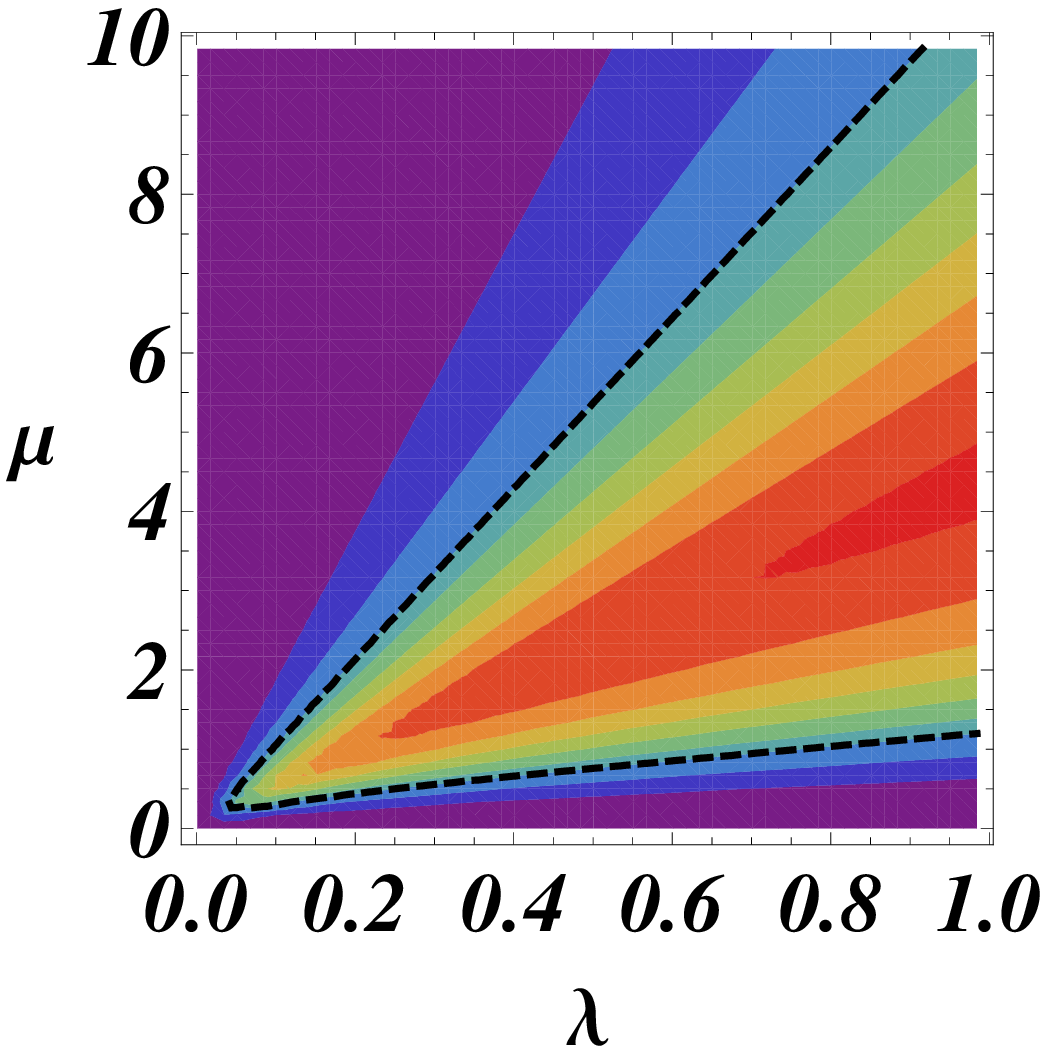}\hspace{.2cm}
\includegraphics[width=1.0cm]{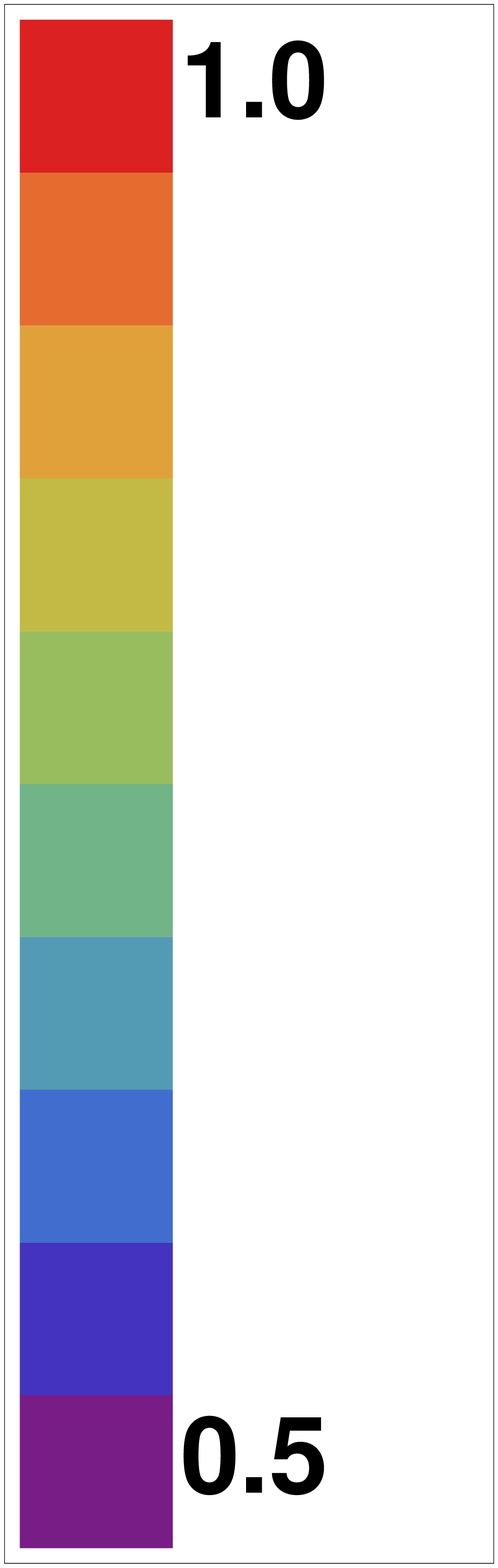}}
\caption{(Color online) Fidelity of teleportation $F_{max}$ in a
$\lambda$--$\mu$ configuration, by exact numerical diagonalization,
for a CCA of $N=12$ cavities as a function of the couplings
$\lambda=J_1/J_b$ and $\mu=J_2/J_b$ at different temperatures
$T/J_b$. Left panel: $T/J_b = 0.005$. Right panel: $T/J_b = 0.01$.
$F_{max}$ varies between $0.5$ (violet) and $1$ (red). Horizontal
dashed line: Classical threshold $F_{max}^c = 2/3$.}
\label{Maxfidelity-1}
\end{figure}

We now proceed to illustrate that CCAs in the $\lambda$--$\mu$
configuration allow for long-distance and high-fidelity quantum
communication in realistic conditions and at moderately high
temperatures. In Fig. \ref{Maxfidelity-1} we report the fidelity of
teleportation $F_{max}$ \cite{Horodecki} as a function of the
reduced couplings $\lambda$ and $\mu$ for different temperatures.
Remarkably, \fig{Maxfidelity-1} shows the existence of a rather high
{\it critical} temperature of teleportation for CCAs realizing a
$\lambda$--$\mu$ spin chain. The region of the physical parameters
compatible with a nonclassical fidelity $F_{max} > 2/3$ is
progressively reduced with increasing temperature, until it
disappears at $T_c \approx 0.13 J_b$. Similar behaviors are observed
for longer CCAs, with $T_{c}$ slowly decreasing with the length of
the array. For instance, for an array of $N=36$ cavities in the
$\lambda$--$\mu$ configuration, the critical temperature of
transition to {\it bona fide} quantum teleportation is $T_c \approx
0.11 J_b$.

A formidable obstacle to the concrete realization of working quantum
teleportation devices is performing the projection over a Bell
state, in order for the sender to teleport a quantum state
faithfully to the receiver. In fact, in the framework of condensed
matter devices there hardly exist quantities, easily available in
current and foreseeable experiments, that admit as eigenstates any
two-qubit Bell states.
To avoid this problem different schemes for quantum information
transfer have been developed, for instance, for what concerns
entanglement transfer, in Ref. \cite{Hartmann-3}.
In the following, we will discuss a simple and concrete scheme for
long-distance, high-fidelity quantum teleportation in
$\lambda$--$\mu$ CCAs that realizes Bell-state projections
indirectly, by matching together free evolutions and local
measurements of easily controllable experimental quantities
\cite{Zheng,Ye,Cardoso}. We first illustrate it in the simplest case
of two cavities at zero temperature, with the first cavity
accessible by the sender and the second one by the receiver. The
sender has access also to a third cavity, the "0" cavity, that is
decoupled from the rest of the chain, and stores the state to be
teleported $\ket{\varphi}=\alpha \ket{\uparrow_0}+\beta
\ket{\downarrow_0}$. The decoupling between the $0-th$ cavity and
the rest of the chain can be achieved removing the degeneracy among
$\ket{0}_0$ and $\ket{1-}_0$ and taking
$|\varepsilon_0-\varepsilon_{1-}| \gg J_0$. The total system is
initially in the state
\begin{equation}\label{groundstate}
\ket{\Psi(0)}=\frac{1}{\sqrt{2}}(\alpha \ket{\uparrow_0}+\beta
\ket{\downarrow_0}) (\ket{\uparrow_1} \ket{\downarrow_2}
+\ket{\downarrow_1}\ket{\uparrow_2}).
\end{equation}
At $t=0$ the state begins to evolve and if $J_0 \gg J_1$ one
has:
\begin{eqnarray}\label{initialstate}
\ket{\Psi(t)}&=&\frac{1}{\sqrt{2}}\left[ \alpha
\ket{\uparrow_0}\ket{\uparrow_1}\ket{\downarrow_2}+\beta
\ket{\downarrow_0}\ket{\downarrow_1}\ket{\uparrow_2} \right.
\\
& & \ket{\uparrow_0}\ket{\downarrow_1} (\alpha \cos(J_0 t)
\ket{\uparrow_2}- i \beta \sin(J_0 t) \ket{\downarrow_2} )\nonumber \\
& &\left. \ket{\downarrow_0}\ket{\uparrow_1} (-i \alpha \sin(J_0 t)
\ket{\uparrow_2}+ \beta \cos(J_0 t) \ket{\downarrow_2}) \right] .
\nonumber
\end{eqnarray}
If at time $t=\pi/(4 J_0)$ Alice measures the local magnetizations
($S_0^z$, $S_1^z$) in the first two cavities, she will find with
probability 1/2 that the teleported state is the image of
$\ket{\varphi}$ under a local rotation. The value $1/2$ for the
probability stems from the fact that any simultaneous eigenstate of
$S_0^z$ and $S_1^z$ can be obtained with equal probability but one
may discard the case in which the total magnetization is equal to
$\pm1$. Realizing a local rotation of $\pm \pi/2$ around $S^z_2$,
with the sign depending on the result of the measurement that the
sender communicates classically to the receiver, the latter recovers
the original state $\ket{\varphi}$ with unit fidelity.
\begin{figure}[t]
\includegraphics[width=6. cm]{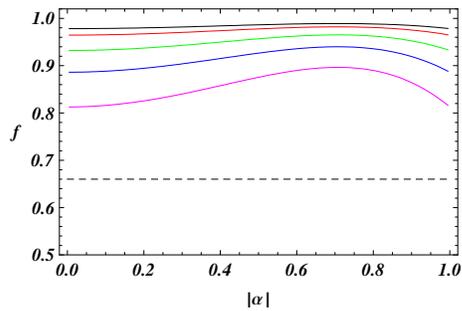}
\caption{(Color online) Fidelity of teleportation $f$ in a
$\lambda$--$\mu$ CCA channel of $N=12$ cavities, for a generic input
$\ket{\varphi}=\alpha \ket{\uparrow_0}+\beta \ket{\downarrow_0}$, as
a function of $|\alpha|$ at different temperatures. Black line: $T =
0.001 J_b $; Red line: $T = 0.003 J_b$; Green line: $T = 0.004 J_b$;
Blue line: $T = 0.005 J_b$; Magenta line: $T = 0.007 J_b$. Here
$\lambda = 0.5$, $\mu = 4.0$, and $\nu = 50$. Horizontal dashed
line: Classical threshold $f_{c}=2/3$.} \label{fidelity}
\end{figure}
The simple protocol described above can be immediately extended to
$\lambda$-$\mu$ CCAs of any size, at finite temperature, and
removing the constraint $J_1 \ll J_0$. By resorting again to exact
diagonalization, in \fig{fidelity} we report the behavior of the
average fidelity of teleportation $f$, as a function of $|\alpha|$
of the state $\ket{\varphi}$, in the case of an array of $N=12$
cavities, with $\nu \equiv J_0/J_b = 50$ and for different
temperatures. Also in the non-ideal case the teleportation protocol
has probability $1/2$ of success. The fidelity depends on the input
state, with a maximum for inputs with $|\alpha|=|\beta|=1/\sqrt{2}$
and a minimum for inputs with $|\alpha|=0.1$. The fidelity remains
above 0.95 for all values of $|\alpha|$ at moderately low
temperatures ($T \simeq 10^{-3}J_{b}$).

\section{Conclusions and outlook}

In conclusion, we have introduced some classes of quantum spin
models defined on open 1-D lattices that are characterized by
non-perturbative GSs with nonvanishing LDE or QLDE. In particular,
end-to-end QLDE is found to be strongly resilient to thermal
decoherence and can be efficiently achieved with a minimal set of
local actions on the end- and near-end couplings in linear open
CCAs. We have showed that QLDE-based CCAs allow for a simple
quasi-deterministic protocol of long-distance, high-fidelity quantum
teleportation that yields a high rate of success without direct Bell
measurements and projections over Bell states. In detail, we have
studied models of $XX$ quantum spin chains with nearest-neighbor
interactions, discussing two types of spatial patterns of the
coupling strengths. In the case of chains with bonds of alternating
strengths, we have shown that the exact GS possesses a nonvanishing
LDE between the end spins of the chain, independently of the size of
the system and asymptotically close to unity (maximal entanglement)
in the limit of exact dimerization. This system is therefore
perfectly suited for \textit{bona fide} long-distance quantum
teleportation with ideal fidelity at zero temperature. However, the
limiting maximal values of the fidelity are obtained at the cost of
introducing an energy gap above the GS that vanishes exponentially
with the size of the system. Therefore, this model is \textit{de
facto} useless for efficient quantum teleportation at finite
temperature. We have then discussed another class of $XX$ open spin
chains with uniform bulk interactions and small end bonds. In this
case, we have shown that for sufficiently small values of the end
couplings, the GS of the system possesses a sizeable QLDE between
the end spins of the chain, that at fixed size of the system is
asymptotically close to unity (maximal entanglement) in the limit of
vanishingly small end bonds. However, QLDE decreases, albeit slowly,
as the size of the system is increased. Models of QLDE can be
improved further by introducing strong near-end bonds that enhance
the value of the end-to-end entanglement ($\lambda$--$\mu$ model).
An interesting feature of QLDE models is that the energy gap above
the GS vanishes only algebraically, as the first power of the
inverse of the size of the system. Therefore, in principle, spin
chains endowed with the QLDE property can be exploited as quantum
channel for teleportation with nonclassical fidelity at finite
temperature.

A relevant issue concerned the behavior of these systems in the
presence of disorder. This question is important especially in view
of possible experimental implementations in which the
couplings can be engineered only within a certain accuracy.
Since naturally the effect of disorder is that of localizing
eigenstates, we expected LDE to be more robust than QLDE against
disorder. In fact, in the LDE scenario the states responsible for the
end-to-end entanglement is already localized at the borders whereas
localization is only approximate in the QLDE case. This conjecture
has in fact been confirmed by thorough exact numerical samplings, that
also show that the relative weight of imperfections that can be
tolerated by the systems, maintaining large and useful values of LDE and QLDE,
is rather high, up to $20\%$, and more. Moreover, for about always
half of the spatial patterns and values of the couplings, disorder
can even enhance the end-to-end entanglement. It is at the
moment unclear how this picture extends to finite temperature.
Another interesting open problem worth further study is to assess
the existence and the possible location of a crossover between
perfect LDE and \textit{prima facie} QLDE.

Demonstrating experimentally (Q)LDE and (Q)LDE-based efficient
long-distance teleportation and state transfer will be a first
crucial preliminary test in order to proceed with integrated devices
of quantum information science, combining atom-optical systems such
as optical lattices or systems of trapped ions with solid-state
based systems, such as those of circuit Cavity Quantum
Electrodynamics.

\acknowledgments This work has been realized in the framework of the
FP7 STREP Project HIP (Hybrid Information Processing). It is a
pleasure to thank Rosario Fazio for some very useful remarks on the
implementation of LDE in optical lattices, and with Lorenzo Campos Venuti,
Cristian Degli Esposti Boschi, and Marco Roncaglia for discussions.
The authors acknowledge financial support from the European Commission of the
European Union under the FP7 STREP Project HIP (Hybrid Information Processing),
Grant Agreement n. 221889, from MIUR under the FARB funds for the years
2007 and 2008, from INFN under Iniziativa Specifica PG 62,
and from CNR-INFM Research Center Coherentia. One of us (F. I.) acknowledges
support from the ISI Foundation for Scientific Interchange.


\begin{thebibliography}{99}

\bibitem{NielsenChuang} M. A. Nielsen and I. L. Chuang, {\it Quantum Computation
and Quantum Information} (Cambridge University Press, Cambridge, UK, 2000).



\bibitem{Gisin} N. Gisin, G. Ribordy, W. Tittel, and H. Zbinden,
Rev. Mod. Phys. \textbf{74}, 145 (2002).

\bibitem{Communication} C. M. Caves and P. D. Drummond, Rev. Mod.
Phys. \textbf{66}, 481 (1994); J. H. Shapiro and O. Hirota (Eds.),
\textit{{} Quantum Communication, Measurements, and Computing}
(Rinton Press, Princeton, N.J., U.S.A., 2003).




\bibitem{Verstraete} F. Verstraete, M. Popp, and J. I. Cirac, Phys.
Rev. Lett. \textbf{92}, 027901 (2004).

\bibitem{Bologna1} L. Campos Venuti, C. Degli Esposti Boschi, and
M. Roncaglia, Phys. Rev. Lett. \textbf{96}, 247206 (2006).

\bibitem{Bologna2} L. Campos Venuti, C. Degli Esposti Boschi, and
M. Roncaglia, Phys. Rev. Lett. {\bf 99}, 060401 (2007).

\bibitem{Salerno1} L. Campos Venuti, S. M. Giampaolo, F. Illuminati,
and P. Zanardi, Phys. Rev. A {\bf 76}, 052328 (2007).

\bibitem{Salerno2} S. M. Giampaolo and F. Illuminati, 
Phys. Rev. A {\bf 80}, 050301(R) (2009).

\bibitem{Bloch} I. Bloch, Nature Phys. {\bf 1}, 23 (2005).

\bibitem{Plenio1} M. J. Hartmann, F. G. S. L. Brand$\tilde{\mathrm{a}}$o, and M. B. Plenio,
Laser \& Photon. Rev. {\bf 2}, 527 (2008).

\bibitem{Sorensen} A. S\"{o}rensen and K. M\"{o}lmer, Phys. Rev. Lett.
{\bf 83}, 2274 (1999).

\bibitem{Duan} L.-M. Duan, E. Demler, and M. D. Lukin, Phys. Rev.
Lett. \textbf{91}, 090402 (2003).

\bibitem{Kuklov} A.~B. Kuklov and B.~V. Svistunov, Phys. Rev. Lett. {\bf 90}, 100401
(2003).

\bibitem{Sanpera} A. Sanpera, A. Kantian, L. Sanchez-Palencia, J. Zakrzewski, and M. Lewenstein
Phys. Rev. Lett. {\bf 93} 040401 (2004).

\bibitem{Hartmann} M. J. Hartmann, F. G. S. L. Brand$\tilde{\mathrm{a}}$o,
and M. B. Plenio, Nature Phys. {\bf 2}, 849 (2006).

\bibitem{Greentree} A. D. Greentree, C. Tahan, J. H. Cole, and L. C. L. Hollenberg, Nature Phys.
{\bf 2}, 856 (2006).

\bibitem{Angelakis} D. G. Angelakis, M. F. Santos, and S. Bose, Phys. Rev. A {\bf 76},
031805(R) (2007).

\bibitem{Rossini} D. Rossini and R. Fazio, Phys. Rev. Lett. {\bf 99}, 186401 (2007).

\bibitem{Hartmann-2} M. J. Hartmann, F. G. S. L. Brand$\tilde{\mathrm{a}}$o,
and M. B. Plenio, Phys. Rev. Lett. {\bf 99}, 160501 (2007).

\bibitem{Lieb} E. Lieb, T. Schultz, and D. Mattis, Ann. Phys. (N.Y.)
\textbf{16}, 407 (1961).

\bibitem{JordanWigner} P. Jordan and E. Wigner, Z. Physik \textbf{47},
631 (1928).

\bibitem{Wootters} W. K. Wootters, Phys. Rev. Lett. \textbf{80},
2245 (1998); S. Hill and W. K. Wootters, Phys. Rev. Lett.
\textbf{78}, 5022 (1997).

\bibitem{Zanardi02} L. Amico, A. Osterloh, F. Plastina, R. Fazio and G. M. Palma, Phys.
Rev. A {\bf 69}, 022304 (2004), P.~Zanardi and X.~Wang, J.~Phys.~ A,
\textbf{35}, 7947 (2002).

\bibitem{Horodecki} M. Horodecki, P. Horodecki, and R. Horodecki, Phys. Rev. A {\bf 60},
1888 (1999).

\bibitem{Badziag} P.~Badziag, M. Horodecki, P. Horodecki, and R.
Horodecki, Phys. Rev. A. \textbf{62}, 012311 (2000).

\bibitem{Woj05} A. Wojcik, T. Luczak, P. Kurzynski, A. Grudka, T.
Gdala, and M. Bednarska, Phys. Rev. A \textbf{72}, 034303 (2005).

\bibitem{Coffman} V. Coffman, J. Kundu, and W. K. Wootters, Phys.
Rev. A, \textbf{61}, 052306 (2000); T. J. Osborne and F. Verstraete,
Phys. Rev. Lett. \textbf{96}, 220503 (2006).

\bibitem{Jaksch} D. Jaksch, C. Bruder J. I. Cirac, C. W. Gardiner, and P. Zoller,
Phys. Rev. Lett. {\bf 81}, 3108 (1998).

\bibitem{Buonsante} P. Buonsante, S. M. Giampaolo, F. Illuminati, V. Penna,
and A. Vezzani, Phys. Rev. Lett. {\bf 100}, 240402 (2008).

\bibitem{Iskin} M. Iskin and C. A. R. Sa de Melo, Phys. Rev. A {\bf 78}, 013607 (2008).

\bibitem{Illuminati} F. Illuminati and A. Albus, Phys. Rev. Lett. {\bf 93}, 090406 (2004).

\bibitem{Fisher} M. P. A. Fisher, P. B. Weichman, G. Grinstein, D. S.
Fisher, Phys. Rev. B {\bf 40}, 546 (1989).

\bibitem{Logneg} See, e. g., M. B. Plenio, Phys. Rev. Lett. {\bf 95}, 090503 (2005),
and references therein.

\bibitem{Das Sarma} C. Zhang, S. L. Rolston, and S. Das Sarma, Phys. Rev. A {\bf 74}, 042316
(2006).

\bibitem{Cho} J. Cho, Phys. Rev. Lett. {\bf 99}, 020502 (2007).

\bibitem{Jaynes} E. T. Jaynes and F. W. Cummings, Proc. IEEE {\bf 51}, 89 (1963).

\bibitem{Hartmann-3} M. J. Hartmann, M. E. Reuter
and M. B. Plenio, New J. Phys. {\bf 8}, 94 (2006).

\bibitem{Zheng} S.-B. Zheng, Phys. Rev. A {\bf 69}, 064302 (2004).

\bibitem{Ye} L. Ye and G-C. Guo, Phys. Rev. A {\bf 70}, 054303 (2004).

\bibitem{Cardoso} W. B. Cardoso, A. T. Avelar, B. Baseia, and N. G. de Almeida, Phys. Rev. A {\bf 72},
045802 (2005).

\end{thebibliography}
\end{document}